\documentclass[sn-basic,sort&compress,icol]{sn-jnl}
\usepackage{graphicx}%
\usepackage[dvipsnames]{xcolor}
\usepackage{multirow}%
\usepackage{amsmath,amssymb,amsfonts}%
\usepackage{amsthm}%
\usepackage{mathrsfs}%
\usepackage[title]{appendix}%
\usepackage{textcomp}%
\usepackage{manyfoot}%
\usepackage{booktabs}%
\usepackage{algorithm}%
\usepackage{algorithmicx}%
\usepackage{algpseudocode}%
\usepackage{listings}%
\usepackage{tikz}%
\tikzset{>=latex}
\usepackage{pgfplots}%
\usepackage{multirow}
\usepackage{bbold}
\usetikzlibrary{calligraphy}
\usepackage{subcaption}
\usepackage{graphicx}
\usepackage{stfloats}

\pgfplotsset{%
 compat=newest, %footnotesize
 tick label style={font=\footnotesize},
 label style={font=\small},
 legend style={font=\small},
 axis x line = center,
 axis y line = center,
 every axis/.style={pin distance=1ex},
 trim axis left
 }

\newcommand{\data}{y}
\newcommand{\Data}{\textbf{y}}
\newcommand{\boldtheta}{{\boldsymbol{\theta}}}
\newcommand{\textbfz}{\textbf{z}}
\newcommand{\Rstat}{\textbf{T}(\Data)}
\newcommand{\Obsrstat}{\textbf{T}_0}
\theoremstyle{thmstyleone}%
\theoremstyle{thmstyletwo}%
\newtheorem{remark}{Remark}%
\theoremstyle{thmstylethree}%
\begin{document}

\title[Article Title]{Insufficient Gibbs Sampling}

\author*[1]{\fnm{Antoine} \sur{Luciano}}\email{antoine.luciano@dauphine.psl.eu}

\author[1,2]{\fnm{Christian P.} \sur{Robert}}\email{xian@ceremade.dauphine.fr}

\author[1]{\fnm{Robin J.} \sur{Ryder}}\email{ryder@ceremade.dauphine.fr}

\affil*[1]{\orgdiv{CEREMADE}, \orgname{Université Paris Dauphine}}

\affil[2]{\orgdiv{Department of Statistics}, \orgname{University of Warwick}}

\abstract{In some applied scenarios, the availability of complete data is restricted, often due to privacy concerns; only aggregated, robust and inefficient statistics derived from the data are made accessible. These robust statistics are not sufficient, but they demonstrate reduced sensitivity to outliers and offer enhanced data protection due to their higher breakdown point. We consider a parametric framework and propose a method to sample from the posterior distribution of parameters conditioned on various robust and inefficient statistics: specifically, the pairs (median, MAD) or (median, IQR), or a collection of quantiles. Our approach leverages a Gibbs sampler and simulates latent augmented data, which facilitates simulation from the posterior distribution of parameters belonging to specific families of distributions.
A by-product of these samples from the joint posterior distribution of parameters and data given the observed statistics is that we can estimate Bayes factors based on observed statistics via bridge sampling. We validate and outline the limitations of the proposed methods through toy examples and an application to real-world income data.}

\maketitle

\section{Introduction}

Traditional statistical methods are sensitive to deviations from Gaussian assumptions, as highlighted by \citet{tuke\data_j_surve\data_1960}. 
Robust statistics are used to safeguard against these deviations; theoretical advancements by \citet{huber_robust_1964} and \citet{hampel_contributions_1968} laid the foundation for robust statistical techniques. 

Robust statistics are also used for privacy reasons. Due to data protection laws, the sharing of sensitive personal data is restricted among businesses and scientific institutions. 
To address this, organizations such as Eurostat and the World Bank often do not release individual-level data $\Data$, but only a robust and insufficient aggregated, multidimensional summary statistic $\Rstat$ instead.
In other cases, observations may be summarized by robust statistics to reduce the impact of outliers or of model misspecification.
This limitation creates a need for statistical methods that can effectively infer parameters from observed robust statistics.
In a Bayesian setting, we might impose a parametric distribution $\left(\mathcal F_{\boldtheta}\right)_{\boldtheta\in\Theta}$ on the original observations, and aim at sampling from the posterior distribution of $\boldtheta$ given robust statistics. 
The posterior distribution is typically intractable, making its simulation challenging and an interesting area of research. Previous studies have employed Approximate Bayesian Computation (ABC) with robust summary statistics, such as the median, Median Absolute Deviation (MAD), or Interquartile Range (IQR) \citep{turner_tutorial_2012,green_bayesian_2015,marin_relevant_2014}. \citet{huang_learning_2023} argue that for ABC or other simulation-based inference methods, robust summary statistics make the pseudo-posterior robust to model misspecification \citep{frazier2020model}. While ABC provides an approach to infer posterior distributions when likelihood evaluations are difficult, these methods only enable simulation from an approximation of the posterior distribution and are less satisfactory than a scheme to sample from the exact posterior. Matching quantiles have also been explored in various contexts \citep{nirwan_bayesian_2022,mcvinish_improving_2012}.

We introduce a method to sample from the joint posterior distribution of the parameters $\boldtheta\in \Theta$ and the full data $\Data\in \mathcal X \subset \mathbb R^N$ given the summary robust and insufficient observed statistic $\Rstat = \Obsrstat$, using a data-augmentation approach within the framework established by \citet{tanner_calculation_1987}.
Since our method samples from the exact posterior, it is an improvement on the existing ABC methods.

For simplicity of notation throughout this paper, the conditioning will be denoted simply as $\cdot \mid \Obsrstat$. We rely on the decomposition:
\begin{equation}
\label{eq:joint1}
 \pi(\boldtheta,\Data\mid \Obsrstat) \propto \pi(\boldtheta\mid \Data)\pi(\Data\mid\Obsrstat).
\end{equation}
and construct a two-step Gibbs sampler which samples alternately 1) from the full conditional $f(\Data\mid \boldtheta,\Obsrstat)$, and 2) from $\pi(\boldtheta\mid\Data,\Obsrstat)$ which, due to information redundancy, is equivalent to the posterior $\pi(\boldtheta\mid\Data)$; this is similar to the strategy outlined in \citet{lewis2021bayesian}. 
Samples of $\boldtheta$ generated via this algorithm marginally conform to the posterior $\pi(\boldtheta\mid \Obsrstat)$. The execution of the second step is direct in scenarios where the distribution belongs to a family allowing for conjugate priors, such as in Gaussian distributions, or it can be implemented through a Metropolis-within-Gibbs step for other families.

We discuss in detail the first step which is typically intractable when $\textbf{T}$ is not sufficient; this represents one of the main contributions of this work. This step involves simulating from the parametric family 
$\left(\mathcal F_\boldtheta\right)_{\boldtheta\in\Theta}$ 
while ensuring the simulated data preserves the observed statistic $\Obsrstat$, i.e., generating $\Data \sim \mathcal F_\boldtheta$ such that $\Data \in \mathcal X_{\Obsrstat} = \{\Data \in \mathcal X \mid \Rstat = \Obsrstat\}$. 

Our methodology is general, and we consider in detail the specific cases where $\textbf{T}$ is a pair of robust location and scale statistics, such as (median, Median Absolute Deviation) and (median, Interquartile Range), as well as cases where $\textbf{T}$ is a collection of empirical quantiles of $\Data$.
Our only assumption on the family of distributions $\mathcal F_{\boldtheta}$ is that we can evaluate pointwise the probability density function $f_{\boldtheta}$ and the cumulative density function $F_{\boldtheta}$.
In this setting, it is in particular possible to sample from a truncated distribution, either directly or by rejection sampling.

In the second part of this paper, we consider
Bayesian model choice based solely on observing sufficient or insufficient summary statistics. This situation is commonly addressed through ABC model choice, which estimates the Bayes factor based on the observed statistic, denoted as $B_{12}^{\textbf{T}}(\Obsrstat)$. This value may not always be convergent for model selection \citep{Robert2011,marin_relevant_2014,didelot}, but when it is relevant, it is approximated using an ABC tolerance $\epsilon>0$. In this paper, we introduce a method that leverages samples $(\boldtheta,\Data)$ from the joint distribution \eqref{eq:joint1} to precisely estimate this value (within the Monte Carlo error) through bridge sampling. This approach allows for a more accurate and direct computation of the Bayes factor, improving upon traditional ABC methods by eliminating the need for tolerance thresholds and providing a framework for model comparison and selection in the case the statistic $\textbf{T}$ is also robust for model choice. 

The manuscript is organized as follows: Section \ref{sec:method} presents our methodological framework for generating augmented data when a sequence of quantiles is observed, detailed in Section \ref{sec:quantile}. We then elaborate on the scenarios of observing the median and interquartile range (Section \ref{sec:iqr}), as well as the particularly interesting case of observing only the median and the Median Absolute Deviation (MAD) of the sample (Section \ref{sec:mad}). Section \ref{sec:BMC} introduces and discusses a strategy for conducting Bayesian model choice under these specific observational settings. The paper concludes with Section \ref{sec:results}, where we provide a thorough examination of numerical findings derived from our proposed methods.

\section{Methodology for sampling from $\Data\mid \Obsrstat$}
\label{sec:method}
\subsection{Quantile case}
\label{sec:quantile}

We first present the case where the observed robust statistic is a set of quantiles. This setting has already been considered in the literature, but our approach uses a different method, which we will extend in later sections to more complex sets of robust statistics.

In this section, we consider the case where we observe a vector of $M\in \mathbb{N}^*$ quantiles of the data $\Data$.
A collection of probabilities $(p_j)_{j=1\ldots M}$ is pre-specified, and we observe $\Obsrstat = (q_j)_{j=1\ldots M}$ where $q_j$ is the empirical $p_j$ quantile of $\Data$.

\citet{akinshin_quantile_2022} also proposed an MCMC method, implemented in STAN (NUTS or HMC versions), to sample from the posterior distribution when only quantiles are observed. However, they treat the observed quantiles as theoretical ones, and thus assume that they observe the collection $\left(F_{\boldtheta}^{(-1)}\left(p_j\right)\right)$. 
This assumption is reasonable when the sample size $N$ is large.
However, observed quantiles are actually calculated differently in most standard software, and this assumption can lead to a bias in the posterior inference of the parameters, especially with small sample sizes $N$. 
Therefore, in this paper, we adopt a different approach by considering the observed quantiles as empirical quantiles obtained from the widely used quantile estimator $Q(\cdot, p)$, as defined in \citet[Definition 7]{hyndman_sample_1996}. 
This estimator is commonly implemented in major statistical software: it is for example the default of the \texttt{quantile()} function in R, the default of the Python function \texttt{numpy.quantile}, the default of the Julia function \texttt{Statistics.quantile!}, and the behavior of the \texttt{PERCENTILE} function in Excel. 
It is defined for $p<1$ by the following formula:
\begin{equation}
Q(\Data,p) = (1-g) \data_{\scriptscriptstyle{\left( i \right)}}+g \data_{\scriptscriptstyle{\left( i + 1\right)}},
\label{eq:quantile}
\end{equation}
where $h = (N-1)p +1$, $i = \lfloor h \rfloor$ (the integer part of $h$), $\data_{(i)}$ is the $i$th order statistic of $\Data$, and $g = h - i$ (the fractional part of $h$). We will later note those variables $h_j, i_j$ and $g_j$ for the observed $p_j$ quantile.

\citet{hyndman_sample_1996} discuss alternative definitions of empirical quantiles, which are available in certain software packages.
Their definitions 1 and 3 provide a purely deterministic framework based on a single order statistic and we do not consider these definitions here. Definitions 2 and 4 to 9 in \citet{hyndman_sample_1996} are similar to our method, since they involve linear combinations of two order statistics. Our methodology can be easily adjusted to accommodate any of these alternative definitions by simply modifying the definitions of $h$ and $g$.

We now develop a computational method to simulate a vector $\Data$ that follows a distribution $\mathcal{F}_{\boldtheta}$ and satisfies the conditions $Q(\Data,p_j) = q_j$ for $j=1,\dots,M$, where $Q$ is the quantile estimator.

\begin{figure*}[t]
\begin{center} 
 \begin{tikzpicture}[scale=1.8]
 \begin{axis}[%
 axis x line=center,
 axis y line=none,
 xmin=0,xmax=1, ymin=0, ymax=3,
 xtick={0.15,0.3,0.7,.85},
 xticklabels={},
 ]
 \end{axis}
 \node [draw=none] at (.5,.35) {$p_1$} ;
 \node [draw=none] at (1.55,0.35) {$p_2-p_1$} ;
 \node [draw=none] at (3.4,0.35) {$\dots$} ;
 \node [draw=none] at (5.3,0.35) {$p_{M}-p_{M-1} $} ;
 \node [draw=none] at (6.35,0.35) {$1-p_M$} ;
 \node [draw=none] at (1,-0.2) {$q_1$} ;
 \node [draw=none] at (2,-0.2) {$q_2$} ;
 \node [draw=none] at (3.4,-0.2) {$\dots$} ;
 \node [draw=none] at (4.8,-0.2) {$q_{M-1}$} ;
 \node [draw=none] at (5.8,-0.2) {$q_M$} ;
 \end{tikzpicture}
\caption{Apportionment of the vector $\Data$ with observed $p_j$ quantiles $q_j$. The values above the axis represent the theoretical proportions of observations contained in these intervals.}
\label{fig:quantile}
\end{center}

\end{figure*}
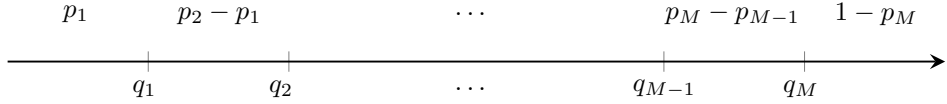

In this scenario, we have complete knowledge of the apportionment of the vector $\Data$ across $M+1$ intervals. The theoretical apportionment is presented in Figure \ref{fig:quantile}.
However, as we consider the empirical quantiles here, these intervals and proportions may slightly vary.

We initialize our vector so that it matches the observed quantiles (see Appendix \ref{sec:init_iqr} for details), then at each iteration we resample the vector $\Data$ while preserving the verified conditions $(Q(\Data,p_1),\dots,Q(\Data,p_M))=(q_1,\dots,q_M)$.
Here, to achieve a better Effective Sample Size on our parameter chains and because it is feasible, we present a method that fully resimulates our augmented data at each iteration. However, a similar approach that only partially updates the data, conditional on the rest of the dataset, can be considered to reduce computational costs when $N$ is large.

We proceed by first simulating the coordinates of $\Data$ that determine the observed quantiles $Q(\Data,p_1),\dots,Q(\Data,p_M)$. Second, we simulate the remaining coordinates of $\Data$ using truncated distributions, ensuring that the correct number of coordinates falls within each zone defined by the previously simulated coordinates. We detail the first step of this process; the second is straightforward.

To simulate these coordinates according to the correct distribution, we must first identify the indexes of the order statistics. From the above definition, we have for $j = 1,\dots,M$:

\begin{itemize}
 \item[-] If $g_j = 0$, we have $Q(\Data,p_j) = \data_{\scriptscriptstyle{(i_j)}}$, which we refer to as ``deterministic", and we denote $i_j$ as its index.
 \item[-] If $g_j \neq 0$, we have $Q(\Data,p_j) = (1-g_j) \data_{\scriptscriptstyle{(i_j)}}+g_j \data_{\scriptscriptstyle{(i_j+1)}}$, which we say is a linear combination of the order statistics with indexes $i_j$ and $i_j+1$. In this case, we sample $\data_{\scriptscriptstyle{(i_j)}}$, and then obtain $\data_{\scriptscriptstyle{(i_j+1)}}$ as a deterministic transformation of $\data_{\scriptscriptstyle{(i_j)}}$ and $q_j$.
\end{itemize}

We denote $J_D = \{j \in \{1,\dots,M\} \mid g_j = 0\}$ and $J_S = \{j \in \{1,\dots,M\} \mid g_j > 0\}$. We have $\{1,\dots,M\} = J_D \cup J_S$. Thus, the quantiles of interest $Q(\Data,p_1),\dots,Q(\Data,p_M)$ are totally determined by the order statistics of indexes in $I = \{i_j \mid j =1,\dots,M\} \cup \{i_j+1 \mid j \in J_S\}$.

\vspace{.2cm}

\begin{remark}
For simplicity of exposition, we assume that for all $j$, $p_{j+1} - p_j \geq \frac{2}{N+1}$. Under this assumption, each order statistic appears at most once in the set of empirical constraints of the form of Equation \ref{eq:quantile}; in other words, we assume that for all $j\in J_S, i_{j}+1 < i_{j+1}$. However, in practice, when $N$ is small, an order statistic may influence two different observed quantiles. This case does not pose any extra mathematical difficulties and has been implemented in the code.
\end{remark}

\vspace{.2cm}

Therefore, we need to simulate the order statistics $(\data_{\scriptscriptstyle{(i_j)}})_{j \in J_S}$; recall that the $(\data_{\scriptscriptstyle{(i_j)}})_{j \in I\setminus J_S}$ are deterministic conditional on the $(\data_{\scriptscriptstyle{(i_j)}})_{j \in J_S}$. Our aim is to simulate from the conditional distribution $(\data_{\scriptscriptstyle{(i_j)}})_{j \in J_S} \mid Q(\Data,p_j)=q_j$ for $j = 1,\dots,M$. To achieve this, we compute the density of this distribution up to a constant, enabling us to launch a Markov chain that targets this distribution using a Metropolis-Rosenbluth-Teller-Hastings (MRTH) kernel.

We begin by considering the joint density of the order statistics vector $I$. The joint probability density function of $M$ statistics of order $(i_1,\dots,i_M)$ from a vector $\Data$ of size $N$ following a distribution $\mathcal{F}_{\boldtheta}$ with density $f_{\boldtheta}$ and cumulative distribution function (cdf) $F_{\boldtheta}$, can be expressed as follows \citep{david_order_2004}:

\begin{equation} \label{eq:order}
    f(\data_1,\dots,\data_M) \propto \prod_{j=1}^M f_{\boldtheta}(\data_j) \prod_{j=0}^M \frac{\left(F_{\boldtheta}(\data_{j+1})-F_{\boldtheta}(\data_j)\right)^{i_{j+1}-i_j-1}}{(i_{j+1}-i_j-1)!}
\end{equation},

\noindent where $\data_0 = -\infty$, $\data_{\scriptscriptstyle{M+1}} = +\infty$, $i_0= 0$, and $i_{\scriptscriptstyle{M+1}}=N+1$. 

To simulate from the joint distribution of $\left(\data_{\scriptscriptstyle{(i_j)}}\right)_{j\in J_S}$ and $\left(Q(\Data,p_j)\right)_{j=1,\dots,M}$, we perform a change of variables denoted as $\phi$. This transformation is injective and continuously differentiable, ensuring that the determinant of its Jacobian is nonzero. The transformation is described below by the second system:
$$\left\{\begin{aligned} 
 q_j &= \data_{\scriptscriptstyle{(i_j)}} &\forall j \in J_D \\
 q_j &= (1-g_j)\data_{\scriptscriptstyle{(i_j)}}+g_j \data_{\scriptscriptstyle{(i_j+1)}} &\forall j \in J_S 
 \end{aligned}\right.\\
 \\
\iff \left\{\begin{aligned}
 \data_{\scriptscriptstyle{(i_j)}} &= q_j &\forall j \in J_D \\
 \data_{\scriptscriptstyle{(i_j+1)}} &= \frac{q_j - \data_{\scriptscriptstyle{(i_j)}}(1-g_j)}{g_j} &\forall j \in J_S 
 \end{aligned}\right.
$$

Finally, as the observed values $q_j$ are fixed, we know that the densities of the joint and conditional distributions are proportional. Hence, we have:

\begin{equation*}
 \begin{split}
 &f_{(\data_{\scriptscriptstyle{(i_j)}})_{j \in J_S}\mid(Q(\Data,p_j))_j=(q_j)_j}(\data_1,\dots,\data_{\lvert J_S\rvert})\\
 &\propto f_{(\data_{\scriptscriptstyle{(i_j)}})_{j \in J_S},(Q(\Data,p_j))_j=(q_j)_j}(\data_1,\dots,\data_{\lvert J_S\rvert},(q_j)_j) \\
 &\propto f_{(i)_{i \in I}}(\phi^{-1}(\data_1,\dots,\data_{\lvert J_S\rvert},q_1,\dots,q_j)).
 \end{split}
\end{equation*}

We have now obtained the conditional density, up to a constant, of the order statistics of interest given the observed quantiles. This enables us to simulate data based on our specified conditions. Therefore, we can construct a Markov chain that targets the desired distribution by employing a MRTH acceptance kernel. In our case, we use a random walk kernel with a variance that can be empirically adjusted. While it is possible to resample all the order statistics simultaneously using a kernel of size $\mathbb{R}^{\lvert J_S \rvert}$, for the purpose of achieving higher acceptance rates, we resample them one by one or in parallel. 

To maximize acceptance, we recommend normalizing the variance of the kernel of $\data_{\scriptscriptstyle{(i_j)}}$ by a constant $\Tilde{c_j}=\text{Var}(\data_{\scriptscriptstyle{(i_j)}}) /(1-g_j)$, assuming that $\Data \sim \mathcal{F}_{\boldtheta}$. Here, we approximate the variance of the order statistics using the formula presented in \citet[p.~120]{baglivo_2005}: Var$(\data_{(i)})\approx \frac{p_i(1-p_i)}{(N+2) f_{\boldtheta}(Q_{\boldtheta}(p_i))^2}$, where $N$ is the sample size, $p_i = \frac{i}{N-1}$, and $f_{\boldtheta}$ and $Q_{\boldtheta}$ are the density and quantile functions of our distribution. This approximation allows us to handle some cases of order statistics with infinite variance as for the Cauchy distribution.

Implementation results for this case on real world data are shown in Section \ref{sec:real_data}.

\subsection{Median and IQR case}
\label{sec:iqr}

We now present a computational method to simulate a vector $\Data$ which follows a distribution $\mathcal{F}_{\boldtheta} $ and verifies the conditions $\text{median}(\Data) = m$ and $\text{IQR}(\Data) = i$ where $m\in \mathbb{R}$ and $i>0$. Here, $\text{median}(\Data)$ is the median of $\Data$, and $\text{IQR}(\Data)$ is the interquartile range of $\Data$. The interquartile range is the difference between the 0.75 quantile, which is the third quartile denoted $Q_3$, and the 0.25 quantile, which is the first quartile denoted $Q_1$, i.e., $\text{IQR}(\Data) = Q(\Data,0.75) - Q(\Data,0.25) = Q_3 - Q_1$. 
This scale estimator has a long history in robust statistics, dating back to the early development of robust estimation techniques. Its resistance to outliers, as quantified by its breakdown point of 25\%, has made it a fundamental tool in robust statistical analysis. Today, the IQR continues to hold a prominent position in robust statistics due to its properties and its ability to summarize the variability of a dataset in a resistant manner. The IQR, being equal to twice the MAD in the case of a symmetric distribution, not only measures the dispersion of the data but also offers a way to capture the asymmetry of the distribution.

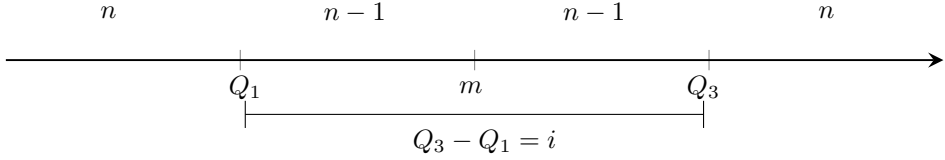
\begin{figure*}[t]
\begin{center} 
 \begin{tikzpicture}[scale=1.8]
 \begin{axis}[%
 axis x line=center,
 axis y line=none,
 xmin=0,xmax=1, ymin=0, ymax=3,
 xtick={0.25,0.5,0.75},
 xticklabels={},
 ]
 \end{axis}
 \node [draw=none] at (.75,.35) {$n$} ;

 \node [draw=none] at (2.55,0.35) {$n-1$} ;

 \node [draw=none] at (4.3,0.35) {$n-1$} ;

 \node [draw=none] at (6,0.35) {$n$} ;

 \node [draw=none] at (1.75,-0.2) {$Q_1$} ;
 \node [draw=none] at (3.4,-0.2) {$m$} ;
 \node [draw=none] at (5.1,-0.2) {$Q_3$} ;
 \draw (1.75,-.4) -- (5.1,-.4);
 \draw (1.75,-.3) -- (1.75,-.5);
 \draw (5.1,-.3) -- (5.1,-.5);
 \node [draw=none] at (3.5,-0.6) {$Q_3 - Q_1 = i$};
 \end{tikzpicture}
\end{center}
\caption{Apportionment of the vector $\Data$ of size $N=4n+1$ with $n\in \mathbb{N}^*$ where median$(\Data)=m$, IQR$(\Data)=Q_3-Q_1=i$}
\label{fig:iqr}

\end{figure*}

This section is linked to the previous section, focusing on the case where $p_1, p_2,$ and $p_3$ take values $0.25, 0.5,$ and $0.75$ respectively. However, in this scenario, we only observe $q_{2}=m$ (the median) and the difference $q_{3} - q_{1}=i$ with $i>0$. Similar to the quantile case, we have knowledge of the distribution of the data vector, which is described in Figure \ref{fig:iqr}.

Once more, we leave the method of initializing the vector $\Data^{0}$ to Appendix \ref{sec:init_iqr}, and we present a method for fully resimulating the augmented data at each iteration while keeping the observed statistics unchanged, i.e., median$(\Data)=m$, IQR$(\Data)=i$. 

In this situation, we can isolate four different cases depending on the value of $(N \mod 4)$. Here, we focus on the specific scenario where $N = 4n+1$ (see Appendix \ref{sec:appendi\data_iqr} for other cases), which simplifies the exposition as no linear interpolation is required to compute the empirical quartiles. Indeed, in this case, the first quartile $Q_1 = \data_{(n+1)}$, the median $Q_2 = m = \data_{(2n+1)}$, and the third quartile $Q_3 = \data_{(3n+1)}$ are all based on a single order statistic. Thus, given that the value $\data_{(2n+1)} = q_2$ is deterministic, we simulate $\data_{(n+1)}$, and then deterministically update $\data_{(3n+1)} = \data_{(n+1)} +i$.
Hence, our goal is to simulate $\data_{(n+1)}$ according to the conditional distribution $\data_{(n+1)} \mid \text{median}(\Data) = m, \text{IQR}(\Data) = i$.

Using the general framework for quantiles described in section \ref{sec:quantile}, we start with the joint distribution of the order statistics $\data_{(n+1)}, \data_{(2n+1)}, \data_{(3n+1)}$ given by Equation \eqref{eq:order}, and apply a change of variables to obtain the joint distribution of the first quartile, the median and the IQR.
As in the previous section, we then use this density in a Metropolis-within-Gibbs step.

The cases where $N\neq 4n+1$ involve more order statistics since the empirical quartiles comprise a linear interpolation, but the same strategy applies. We give details in Appendix \ref{sec:appendi\data_iqr}.
\subsection{Median and MAD case}
\label{sec:mad}

We now focus on the most intriguing scenario explored in this paper, where we are provided with the median (a robust statistic for location) and the MAD (a robust statistic for scale).

Recall that the median is the 0.5 quantile of our sample $\Data$ and is defined as follows:
$\text{median(}\Data\text{)} = \left\{
 \begin{array}{ll}
  \data_{(n)} \mbox{, if } N=2n+1 \\
  \left(\data_{(n)}+\data_{(n+1)}\right)/2\mbox{, if } N=2n
 \end{array}
\right.$ where $\data_{(i)}$ denotes the $i$th order statistic of $\Data$. 

The Median Absolute Deviation (MAD) is a measure of statistical dispersion that is commonly used as a robust alternative to the standard deviation. This statistic was first promoted by \citet{hampel_influence_1974}, who attributed it to \citet{gauss_bestimmung_1880}. For a sample $\Data = (\data_1,\dots, \data_N)$ of i.i.d. random variables, the MAD is defined as: 
$$\text{MAD(}\Data\text{)} = \text{median(}\lvert \Data - \text{median(}\Data\text{)}\rvert\text{).}$$
Let $\sigma$ be the true standard deviation of the data generating distribution. For certain families of distribution, a family-specific constant $c$ is known such that MAD$(\data_1,\ldots,\data_N)\xrightarrow[n\to\infty]{\mathbb P}c\sigma$. Some papers thus refer instead to the normalized MAD, defined as $\mathrm{MAD}(\Data) / c$, which provides a consistent estimator of the standard deviation.

Despite their poor statistical efficiencies (respectively 63.6\% and 36.7\%), a key similarity between the median and the MAD that makes them popular is their breakdown point. Both the median and the MAD have a breakdown point of 50\%, meaning that half of the observations in the dataset can be contaminated without significantly impacting their estimates. This high breakdown point ensures the robustness of these estimators in the presence of outliers and underscores their usefulness in robust statistical analysis.

Since the median and MAD are based on order statistics, the cases where $\Data$ has an even or odd size exhibit distinct characteristics. Here, we focus on the simpler case where $N$ is odd i.e $N=2n+1$ with $n\in \mathbb{N}^*$; we relegate the even case to Appendix \ref{sec:even}. We denote $\text{median}(\Data) = m$ and $\text{MAD}(\Data) = s$ respectively, where $m \in \mathbb{R}$ and $s > 0$. In this scenario, since $\mathcal F_{\boldtheta}$ is continuous, the median is necessarily one of the coordinates of the vector $\Data$, denoted as $\data_i = m$ for some $i \in \{1,\dots,N\}$. 
Additionally, there exists another coordinate $j \in \{1,\dots,N\}$ (if $N>1$) that determines the MAD, such that $\lvert \data_j - m \rvert = s$; we denote the corresponding observation as $\data_{\scriptscriptstyle\mathrm{MAD}}$. 
Note that $\data_{\scriptscriptstyle\mathrm{MAD}}$ can only take two values: $\data_{\scriptscriptstyle\mathrm{MAD}}\in\{m-s,m+s\}$.
We introduce the indicator variable $\delta = \mathbb{1}_{\data_{\scriptscriptstyle{\mathrm{MAD}}} = m+s}$ to capture the location of this second coordinate. We can partition the data into four intervals:

\begin{itemize}
 \item[-] $Z_1 = (-\infty,m-s)$
 \item[-] $Z_2 = (m-s,m)$
 \item[-] $Z_3 = (m,m+s)$
 \item[-] $Z_4 = (m+s,+\infty)$.
\end{itemize}

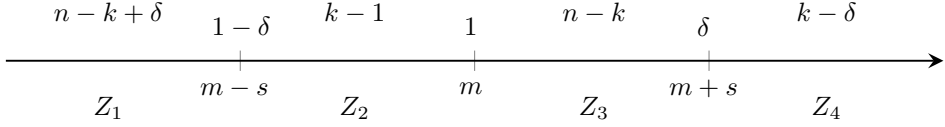
\begin{figure*}[!h]
\begin{center} 
 \begin{tikzpicture}[scale=1.8]
 \begin{axis}[%
 axis x line=center,
 axis y line=none,
 xmin=0,xmax=1, ymin=0, ymax=3,
 xtick={0.25,0.5,0.75},
 xticklabels={},
 ]
 \end{axis}
 \node [draw=none] at (.75,.35) {$n-k+\delta$} ;
 \node [draw=none] at (.75, -.35) {$Z_1$} ;

 \node [draw=none] at (2.55,0.35) {$k -1$} ;
 \node [draw=none] at (2.55, -.35) {$Z_2$} ;

 \node [draw=none] at (4.3,0.35) {$n-k$} ;
 \node [draw=none] at (4.3, -.35) {$Z_3$} ;

 \node [draw=none] at (6,0.35) {$k-\delta$} ;
 \node [draw=none] at (6, -.35) {$Z_4$} ;

 \node [draw=none] at (1.67,-0.2) {$m-s$} ;
 \node [draw=none] at (1.72,0.25) {$1-\delta$} ;
 \node [draw=none] at (3.4,-0.2) {$m$} ;
 \node [draw=none] at (3.4,0.25) {1};
 \node [draw=none] at (5.1,0.25) {$\delta$};
 \node [draw=none] at (5.1,-0.2) {$m+s$} ;
 \end{tikzpicture}
 \caption{Apportionment of the vector $\Data$ when $N=2n+1$ with $n\in \mathbb{N}^*$, median$(\Data)=m$ and MAD$(\Data)=s$.}
 \label{fig:mad}

\end{center}
\end{figure*}

These intervals are represented in Figure \ref{fig:mad}.
The constraint on the median implies that $\lvert Z_1 \cup Z_2 \cup \{\data_{\scriptscriptstyle\mathrm{MAD}}\}\rvert = \lvert Z_3 \cup Z_4 \cup \{\data_{\scriptscriptstyle\mathrm{MAD}}\} \rvert= n$. 
Moreover, the MAD requires that half of the data fall within the interval $(m-s, m+s)$ and the remaining half outside of this interval, so we have $\lvert Z_2\cup Z_3 \cup \{m\} \rvert = \lvert Z_1 \cup Z_4 \rvert = n$. %(the $n$th points are respectively at $m$ and $\data_{\scriptscriptstyle\mathrm{MAD}}$).
Let $k=\sum_i \mathbb 1_{\data_i\geq m+s}\in \{1,\dots,n\}$. Given the values of $\delta$ and $k$, the apportionment of the observations between the four zones is fixed, as shown in in Figure \ref{fig:mad}: $|Z_1|=n-k+\delta$, $|Z_2|=k-1$, $|Z_3|=n-k$ and $|Z_4|=k-\delta$.

The initialization step for $\Data^0$ that meets the constraints of the observed median and MAD is once again left to Appendix \ref{sec:init_mad} and is very similar to the interquartile range case. However, the proposed method for updating the vector $\Data^t$ at step $t$ is this time completely different from the other cases presented previously. Indeed, note that the allocation of observations among the four zones, which we have previously defined, is not fixed and can vary between iterations. Specifically, the value of $k$ (which controls the allocation of observations between $Z_1\cup Z_3$ and $Z_2\cup Z_4$) can range from $1$ to $n$% (where $n \in \mathbb{N}^*$ such that $N=2n+1$)
, and the value of $\delta$ can take values $\{0,1\}$. Since to our knowledge, the distribution of $k$ and $\delta$ is unknown, performing a total resimulation of the data becomes unfeasible.

Let $\data_{-i}$, respectively $\data_{-ij}$, be the vector of all coordinates of $\Data$ except coordinate $i$, respectively except coordinates $i$ and $j$. 
A standard Gibbs strategy would be to cycle through the indexes, updating in turn each $\data_i$ conditionally on $\boldtheta$, $\data_{-i}$ and the constraints $m$ and $s$.
This strategy does not adequately explore the full posterior. Indeed, the distribution of $\data_i|\boldtheta,\data_{-i}, m,s$ takes values only in the zone that $\data_i$ belongs to. With such a strategy, the values of $k$ and $\delta$ would never change.
Instead, we must update two coordinates at a time: we will draw randomly two indexes $i$ and $j$ and sample from the joint conditional of $\data_i,\data_j|\boldtheta, \data_{-ij},m,s$.
These joint conditionals are tractable, and we show in Appendix \ref{sec:ergodic} that this produces an ergodic Markov Chain so that the MCMC will explore the full posterior.

The algorithm to generate new values $\Tilde \data_i, \Tilde \data_j$ from this joint conditional distribution is described as follows:
 
\vspace{.2cm}

\begin{enumerate}
 \item If $(\data_i, \data_j) = (m, \data_{\scriptscriptstyle\mathrm{MAD}})$, we must keep their values unchanged: $(\Tilde{\data_i}, \Tilde{\data_j}) = (\data_i, \data_j)$.
 
 \item If $\data_i = m$, we perform the following steps:
 \begin{itemize}
 \item[-] $\Tilde{\data_j}$ is sampled from the distribution $\mathcal{F}_{\boldtheta}$ truncated to the zone to which $\data_j$ belongs.% with the constraint that it belongs to the zone determined by the function $\textrm{Zone}(\data_i)$.
 \item[-] $\Tilde{\data_i}$ remains unchanged: $\Tilde{\data_i} = \data_i=m$.
 \end{itemize}
 
 \item If $\data_j = \data_{\scriptscriptstyle\mathrm{MAD}}$, we further consider two cases:
 \begin{enumerate}
 \item If $(\data_i-m)(\data_j-m)>0$, indicating that both $\data_i$ and $\data_j$ are on the same side of the median, we perform the following steps:
 \begin{itemize}
  \item[-] $\Tilde{\data_i}$ is resampled from the distribution $\mathcal{F}_{\boldtheta}$ truncated to the zone to which $\data_i$ belongs. %with the constraint that it belongs to the zone determined by the function $\textrm{Zone}(\data_i)$.
  \item[-] $\Tilde{\data_j}$ remains unchanged: $\Tilde{\data_j} = \data_j$.
 \end{itemize}
 
 \item If $(\data_i-m)(\data_j-m)\leq 0$, indicating that $\data_i$ and $\data_j$ are on different sides of the median, we perform the following steps:
 \begin{itemize}
  \item[-] $\Tilde{\data_i}$ is sampled from the distribution $\mathcal{F}_{\boldtheta}$ in the union of the zones which $\data_i$ belongs and its ``symmetric":$\left\{
 \begin{array}{ll}
  Z_1\cup Z_4 & \mbox{if } \data_i\in Z_1\cup Z_4 \\
  Z_2 \cup Z_3 & \mbox{if } \data_i\in Z_2 \cup Z_3\\
 \end{array}
\right.
$
  \item[-] $\Tilde{\data_j}$ is determined based on the value of $\Tilde{\data_i}$ to maintain the same number of observations on either side of the median $m$:
  \begin{itemize}
  \item[-] If $\Tilde{\data_i} > m$, then $\Tilde{\data_j} = m - s$.
  \item[-] Otherwise, $\Tilde{\data_j} = m + s$.
  \end{itemize}
  Note that the value of $\delta$ may change.
 \end{itemize}
 \end{enumerate}
 
 \item If none of the above conditions are met, indicating that $\data_i$ and $\data_j$ are neither of $m$ or $\data_{\scriptscriptstyle\mathrm{MAD}}$, we further consider two sub-cases. Assume without loss of generality that $\data_i<\data_j$.
 \begin{enumerate}
 \item If either ($\data_i\in Z_1$ and $\data_j\in Z_3$) or ($\data_i\in Z_2$ and $\data_j\in Z_4$), then they can switch together in the other couple of zones (and then the apportionment indicator $k$ changes). We perform the following steps:
 \begin{itemize}
  \item[-] $\Tilde{\data_i}$ is sampled from the distribution $\mathcal{F}_{\boldtheta}$ (without truncation). 
  \item[-] $\Tilde{\data_j}$ is sampled from the distribution $\mathcal{F}_{\boldtheta}$ truncated to the ``complementary" zone: $\left\{
 \begin{array}{ll}
  Z_3 & \mbox{if } \Tilde \data_i\in Z_1 \\
  Z_4 & \mbox{if } \Tilde \data_i\in Z_2 \\
  Z_1 & \mbox{if } \Tilde \data_i\in Z_3 \\
  Z_2 & \mbox{if } \Tilde \data_i\in Z_4 \\
 \end{array}
\right.
$
 \end{itemize}
 
 \item Otherwise, when the zones of $\data_i$ and $\data_j$ are not ``complementary", they are each sampled from the distribution $\mathcal{F}_{\boldtheta}$ truncated to their respective zones:
 \begin{itemize}
  \item[-] $\Tilde{\data_i}$ is sampled from the distribution $\mathcal{F}_{\boldtheta}$ truncated to the zone to which $\data_i$ belongs.
  \item[-] $\Tilde{\data_j}$ is sampled from the distribution $\mathcal{F}_{\boldtheta}$ truncated to the zone to which $\data_j$ belongs.
 \end{itemize}
 \end{enumerate}
\end{enumerate}

Finally, we set $(\data_i, \data_j) = (\Tilde{\data_i}, \Tilde{\data_j})$ to update the values of the selected coordinates.

It can be checked that in each case, the median and MAD conditions are preserved throughout the resampling process. This method allows us to have an ergodic Markov chain on the space of latent variables $\Data$ that satisfy these conditions (proof in Appendix section \ref{sec:ergodic}). The case where $N$ is even follows the same ideas but requires slightly different updates, which are described in Appendix \ref{sec:even}.

Numerical results for this situation are presented in Section \ref{sec:gaussian}.

\section{Bayesian model choice}
\label{sec:BMC}

\subsection{Introduction to Bayesian model choice}

A key contribution of Bayesian data analysis is that it provides a probabilistic way to quantify the evidence that data provide in support of one model. The marginal likelihood, defined as the normalizing constant in the posterior distributions is the natural and standard Bayesian tool for model comparison:
$$w(\Data) = \int_{\boldsymbol \Theta} \pi(\boldtheta) f(\Data\mid\boldtheta)d\boldtheta.$$

% Bayesian model choice proceeds by creating a probability structure across $m$ models (or likelihoods). It introduces the model indicator $\{1,\dots,m\}$ as an extra unknown parameter, associated with its prior distribution, $\pi(\mathcal M_i)$ $(i = 1, \dots , m)$, while the prior distribution on the parameter is conditional on the model $\mathcal M_i$, denoted by $\pi_i(\boldtheta_i)$ and defined on the parameter space $\Theta_i$. In this setting, a natural Bayes procedure associated with a prior distribution $\pi$ is to consider the posterior probability of $\mathcal M_i$:
% $$p(\mathcal M_i\mid \Data) = \frac{\pi(\mathcal M_i) w_i(\Data)}{\sum_k \pi(\mathcal M_k) w_k(\Data)}$$
% where $w_k(\Data)$ denotes the marginal likelihood for model $k$. 

When comparing just two models, $\mathcal M_1$ and $\mathcal M_2$, with their likelihood functions denoted by $f_1(\Data\mid\boldtheta_1)$ and $f_2(\Data\mid\boldtheta_2)$, respectively, it is usual to evaluate the odds of one model against the other. Assuming that both models have equal prior probabilities, that is $\pi(\mathcal M_1) = \pi(\mathcal M_2)$, this comparison simplifies to the calculation of the Bayes factor as introduced by \cite{jeffreys1939theory}:

$$B_{12} = \frac{w_1(\Data)}{w_2(\Data)} = \frac{\int_{\Theta_1} \pi_1(\boldtheta_1) f_1(\Data\mid\boldtheta_1)d\boldtheta_2}{\int_{\Theta_2} \pi_2(\boldtheta_2) f_2(\Data\mid\boldtheta_2)d\boldtheta_2}.$$

In most cases, however, marginal likelihoods are not analytically tractable and must be approximated using Monte Carlo methods like Importance Sampling. Among these, bridge sampling emerges as a particularly effective method for estimating normalizing constants, as highlighted in existing literature.

\subsection{Bridge sampling}
\label{sec:bridge}

Bridge sampling, as introduced by \citet{mengfong} and further developed by \citet{gelmanmeng}, serves as a robust method effective in high-dimensional parameter spaces for the estimation of normalizing constants. Unlike simpler methods that draw samples from only one distribution, bridge sampling uses samples from two distinct distributions. Initially, as described by \citet{mengfong}, it was applied to compute the ratio between two normalizing constants, such as the Bayes factor $B_{12}(\Data)$, where it leverages the posteriors of both models under comparison as its distributions.

When comparing two models within the same parameter space $\Theta$, modulo a reparameterization of both distributions in terms of parameters with identical interpretation, it is usually the case that only the unnormalized posterior densities $\tilde \pi_1(\boldtheta\mid\Data) = \pi_1(\boldtheta) f_1(\Data\mid \boldtheta)$ and $\tilde \pi_2(\boldtheta\mid\Data) = \pi_2(\boldtheta) f_2(\Data\mid \boldtheta)$ are available. In such cases, exploiting the "overlap" between the functions $\tilde \pi_1$ and $\tilde \pi_2$ and using an everywhere positive function $h$ referred to as the \textit{bridge function} leads to the following approximation of the Bayes factor:
\begin{equation*}
 B_{12}(\Data) = \frac{\mathbb E_{\pi_2(\boldtheta\mid\Data)} [\tilde \pi_1(\boldtheta\mid \Data) h(\boldtheta)]}{\mathbb E_{\pi_1(\boldtheta\mid \Data)} [\tilde \pi_2(\boldtheta\mid\Data) h(\boldtheta) ]}\approx \frac{ \sum_{j=1}^T \tilde \pi_1(\boldtheta_{2,j}\mid\Data) h(\boldtheta_{2,j})}{ \sum_{i=1}^T \tilde \pi_2(\boldtheta_{1,i}\mid\Data) h(\boldtheta_{1,i})} = \widehat B_{12}(\Data),
\label{eq:bridge_mc}
\end{equation*}
where $(\boldtheta_{1,1},\dots,\boldtheta_{1,T}) \sim \pi_1(\boldtheta\mid\Data)$ and $(\boldtheta_{2,1},\dots,\boldtheta_{2,T}) \sim \pi_2(\boldtheta\mid\Data)$.

By aiming for a function $h$ which is an optimal "bridge" between the two posteriors \citep{mengfong}, bridge sampling facilitates the estimation of the Bayes factor through a recursive process that refines an initial estimate of the Bayes factor, denoted as $\widehat{B}_{12}(\Data)^{(0)}$, towards convergence. The approximation process for each subsequent iteration, labeled $k+1$, is given by:

\begin{equation*}
\widehat{B}_{12}(\Data)^{(k+1)} = \frac{ \sum_{j=1}^T \frac{l_{2,j}}{l_{2,j} + \widehat{B}_{12}(\Data)^{(k)}}}{ \sum_{i=1}^T \frac{1}{l_{1,i} +\widehat{B}_{12}(\Data)^{(k)}}},
\label{eq:bridge}
 \end{equation*}
where $l_{1,i} = \frac{\tilde \pi_1(\boldtheta_{1,i}\mid \Data)}{\tilde \pi_2(\boldtheta_{1,i}\mid \Data)}$ and $l_{2,j} = \frac{\tilde \pi_1(\boldtheta_{2,j}\mid \Data)}{\tilde \pi_2(\boldtheta_{2,j}\mid \Data)}$. It does not require further simulations and converges in a few iterations to a fixed point \citep{gelmanmeng}.

\subsection{Bayesian model choice with insufficient statistics}

However, the framework of this paper differs from the above in that the data $\Data$ are not available as such but only through an insufficient statistic $\Rstat = \Obsrstat$. In this setting, a logical step would be to perform ABC model choice based on the observed statistic $\Obsrstat$ as summary statistic.
\citet{grelaud} showed that ABC model choice with zero tolerance estimates the following Bayes factor:
\begin{equation}
 B_{12}^\textbf{T} (\Obsrstat) = \frac{\int_{\Theta_1} \pi_1(\boldtheta_1) g_1^\textbf{T}(\Obsrstat\mid\boldtheta_1)d\boldtheta_2}{\int_{\Theta_2} \pi_2(\boldtheta_2) g_2^\textbf{T}(\Obsrstat\mid\boldtheta_2)d\boldtheta_2},
 \label{eq:suff}
\end{equation}
where $g_i^\textbf{T}$ denotes the density of $\Rstat$ when $\Data\sim f_i$.

This value most often differs from the Bayes factor $B_{12}$ based on the whole data $\Data$. As discussed in \citet{didelot} and \citet{Robert2011}, in the specific case when the statistic $\Rstat$ is sufficient for both model $\mathcal M_1$ and model $\mathcal M_2$, the difference between both Bayes factors can be expressed as
$$B_{12}(\Data)= \frac{g_1(\Data)}{g_2(\Data)}B^\textbf{T}_{12}(\Rstat),$$
where the ratio of the $g_i(\Data)$s often behaves like a likelihood ratio. This ratio is equal to 1 solely when the statistic $\textbf{T}$ is furthermore sufficient across models $\mathcal M_1$ and $\mathcal M_2$. It occurs in the special case of Gibbs random fields \citep{grelaud}.

\citet{relevant} demonstrate further that the ranges of the means of the summary statistic $\Rstat$ under both models being compared must differ for the convergence of the Bayes factor based on $\Rstat$.

In the event that the Bayes factor based on the observed statistic $\Rstat=\Obsrstat$, denoted as $B_{12}^\textbf{T}(\Obsrstat)$, is a suitable tool for model comparison, it still needs to be approximated, which up until now could only be achieved through ABC model choice. This resulted in an incompressible error associated with the ABC tolerance level. We now demonstrate how to accurately estimate this value using bridge sampling.

\subsection{Insufficient bridge sampling}
\label{sec:InsBridge}

In the previous section, we introduced a completion technique enabling the exact sampling of parameters and augmented data from the joint posterior distribution $\pi(\boldtheta,\Data\mid\Obsrstat)$. The formulation of this distribution, which initially appears intractable, can be presented in an unormalized format as follows:
\begin{equation}
\label{eq:joint}
\begin{split}
 \pi_i(\boldtheta,\Data\mid \Obsrstat) & \propto \pi(\boldtheta) f_i (\Data\mid \boldtheta) \mathbb I_{\{\Rstat=\Obsrstat\}}\\
 & = \tilde \pi_i(\boldtheta,\Data\mid \Obsrstat).
\end{split}
\end{equation}
Note that since the $\Data$'s are simulated to make the simulated statistic match the observed statistic, the indicator is always equal to $1$ in \eqref{eq:joint}. Therefore, we can rewrite this unnormalized joint posterior distribution as the conventional unnormalized posterior distribution $\tilde \pi_i(\boldtheta\mid \Data) = \pi_i(\boldtheta) f_i(\Data\mid \boldtheta)$.
Given that sampling from this distribution is feasible and its unconstrained version is manageable, the bridge sampling methodology can thus be invoked to compute $B_{12}^\textbf{T}(\Obsrstat)$. Following similar steps as in Section \ref{sec:bridge}, we derive that

\begin{equation}
\widehat{B}^\textbf{T}_{12}(\Obsrstat)^{(k+1)} = \frac{ \sum_{j=1}^T \frac{l_{2,j}}{l_{2,j} + \widehat{B}^\textbf{T}_{12}(\Data)^{(k)}}}{ \sum_{i=1}^T \frac{1}{l_{1,i} +\widehat{B}^\textbf{T}_{12}(\Data)^{(k)}}},
\label{eq:bridgeT}
 \end{equation}
where
$$l_{1,i} = \frac{\tilde \pi_1(\boldtheta_{1,i},\Data_{1,i}\mid \Obsrstat)}{\tilde \pi_2(\boldtheta_{1,i},\Data_{1,i}\mid \Obsrstat)} = \frac{\tilde \pi_1(\boldtheta_{1,i}\mid \Data_{1,i})}{\tilde \pi_2(\boldtheta_{1,i}\mid \Data_{1,i})}$$ 
and 
$$l_{2,j} = \frac{\tilde \pi_1(\boldtheta_{2,j},\Data_{2,j}\mid \Obsrstat)}{\tilde \pi_2(\boldtheta_{2,j}, \Data_{2,j}\mid \Obsrstat)}= \frac{\tilde \pi_1(\boldtheta_{2,j}\mid \Data_{2,j})}{\tilde \pi_2(\boldtheta_{2,j}\mid \Data_{2,j})}$$
converges to a fixed point approximating the Bayes factor attached to the insufficient statistic.

This novel version of bridge sampling provides a consistent approximation of $B_{12}^\textbf{T}(\Rstat)$. 

Therefore, it allows for Bayesian model choice based solely on the observed summary statistic $\Rstat$, under the provision that the statistic discriminates between the considered models. 

However, we must stress that, from a practical perspective, this estimator is only of interest when the sample size $N$ is not too large. Specifically, incorporating augmented data as latent variables alongside the initial parameters has a negative impact on the variance of the $(l_{1,i})$ and $(l_{2,j})$ in \eqref{eq:bridgeT} and therefore the convergence of the Monte Carlo approximation.

\section{Numerical results and discussion}
\label{sec:results}

\begin{figure*}[t!]
 \centering
 \includegraphics[width=1\textwidth]{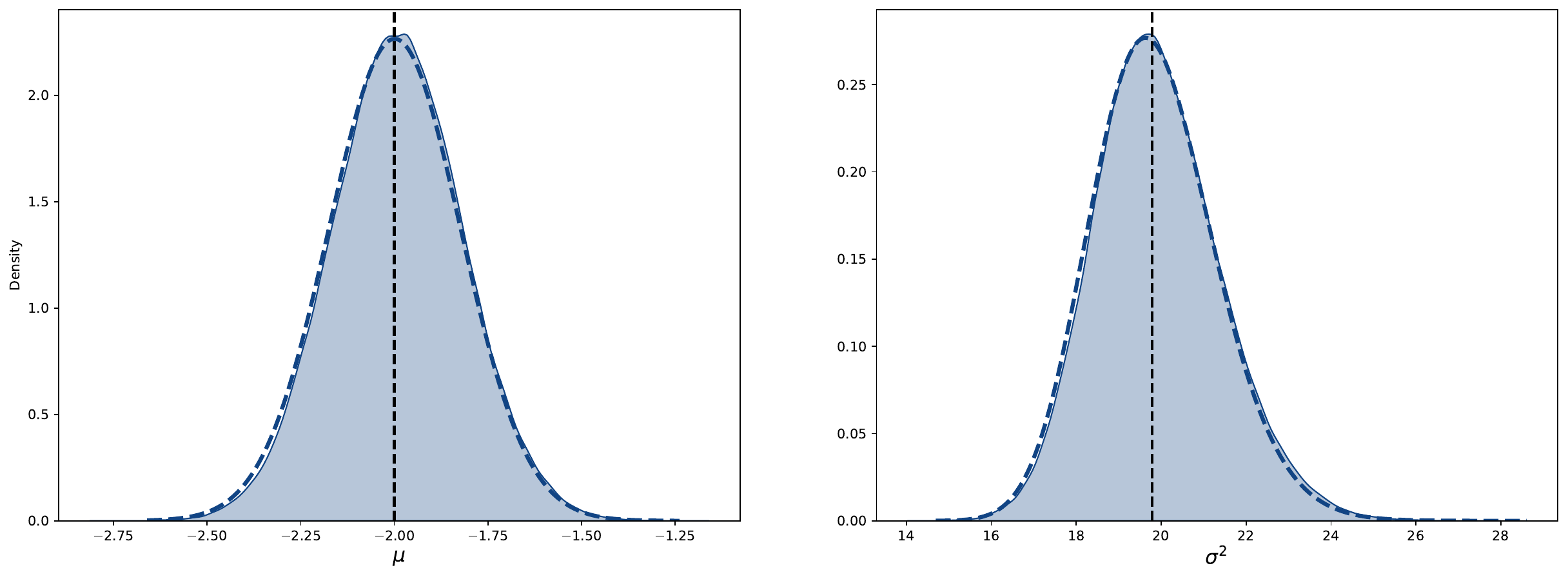}
 
 \caption{Posterior distribution of $\mu$ and $\sigma^2$ for a Normal sample of size $N=1000$, given $\Obsrstat = \Rstat = \left(\text{median}(\Data),\text{MAD}(\Data)\right) = (-2,3)$ obtained by Insufficient Gibbs Sampling (filled curve) and approximated with $\Tilde{\pi}$ (dashed blue line). The theoretical estimands $m$ and $(c \cdot s)^2$ are shown as black dashed lines.}
 \label{fig:gauss_approx}
\end{figure*}

In this section, we discuss the numerical results obtained from the different methods we presented above. 

\subsection{Posterior approximation in the Gaussian case} \label{sec:gaussian}
We initially focus on the Gaussian distribution with conjugate priors distributions for the mean and variance parameters: the Normal-Inverse Gamma distribution (abbreviated into NIG below). This provides a convenient and analytically tractable framework for straightforward sampling from the posterior of the parameters in the second step of the Gibbs Sampler.

In the Gaussian case, the asymptotic efficiencies of the empirical median, empirical MAD, and empirical IQR estimators have been well-studied in the frequentist framework \citep{rousseeuw_alternatives_1993}. The empirical median has an asymptotic efficiency of $\text{eff}_{\text{med}}=\frac{2}{\pi} \approx 0.637$; the empirical MAD has an asymptotic efficiency $\text{eff}_{\scriptscriptstyle\mathrm{MAD}}\approx 0.3675$ \citep{akinshin_quantile_2022}.
These efficiency values measure the relative accuracy of these estimators compared to the conventional estimators (empirical mean and standard deviation in this instance) as the sample size tends to infinity.

Under the prior $\mu, \sigma^2 \sim \text{NIG}(\mu_0, \tau, \alpha, \beta)$ where $\mu \in \mathbb{R}, \tau,\alpha,\beta>0$ are the hyperparameters, the posterior distribution given $\Data$ is known in closed-form: $\mu, \sigma^2 \mid \Data \sim \text{NIG}(M, C, A, B)$ where

\begin{equation}\label{eq:post}
\begin{gathered}
M = \frac{\nu \mu_0 + N \bar{\Data}}{\nu + N}\\
C = \nu + N,\quad A = \alpha + \frac{N}{2} \\
B = \beta + \frac{1}{2}(N S^2 + \frac{N \nu}{\nu + N}(\bar{\Data} - \mu_0)^2).
\end{gathered}
\end{equation}
with $\bar x$ the empirical mean and $S^2$ the empirical variance.

Our algorithm allows us to obtain a sample from the posterior distribution $\pi(\mu, \sigma^2 \mid \text{median}(\Data), \text{MAD}(\Data))$, which is displayed in Figure \ref{fig:gauss_approx}.

Our numerical results give a sample from the exact posterior, and we observe that there exists a high-quality approximation to this posterior when $N$ is large. Returning to Equation \ref{eq:post}, we can replace $\bar \Data$ and $S^2$ by their estimators based on the median and MAD: $\bar \Data\approx m$, $S\approx c\cdot s$ with $c = 1 /\Phi^{-1}(.75)\approx 1.4826$. We also replace the values of $N$ by multiplying by the asymptotic efficiencies of these estimators $N_{\text{med}} = \text{eff}_{\text{med}} \cdot N$ and $N_{\scriptscriptstyle\mathrm{MAD}} = \text{eff}_{\scriptscriptstyle\mathrm{MAD}} \cdot N$.
Figure \ref{fig:gauss_approx} shows that our posterior of interest $\pi(\mu, \sigma^2 \mid \text{median}(\Data), \text{MAD}(\Data))$ is well approximated by the distribution $ \text{NIG}(\Tilde{M},\Tilde{C}, \Tilde{A},\Tilde{B})$ where:

\begin{equation*}\label{eq:post2}
\begin{gathered}
\Tilde{M} = \frac{\nu \mu_0 + N_{\text{med}} \cdot m}{\nu + N_{\text{med}}},\\
\Tilde{C} = \nu + N_{\text{med}}, \quad \Tilde{A} = \alpha + \frac{N_{\scriptscriptstyle\mathrm{MAD}}}{2}\\
\Tilde{B} = \beta + \frac{1}{2}\left(N_{\scriptscriptstyle\mathrm{MAD}} (c s)^2 + \frac{N_{\scriptscriptstyle\mathrm{MAD}} \nu}{\nu + N_{\scriptscriptstyle\mathrm{MAD}}} (m - \mu_0)^2 \right).
\end{gathered}
\end{equation*}

To our knowledge, this high-quality approximation, which can be easily sampled from, was not previously known. Our numerical results apply only to the Gaussian case with observed median and MAD; we leave to future work the question of whether similar results hold for other distribution and robust statistics with known asymptotic efficiency.

\subsection{ABC comparison with the Cauchy distribution}
\begin{figure*}[t!]
 \centering
 \includegraphics[width=1\textwidth]{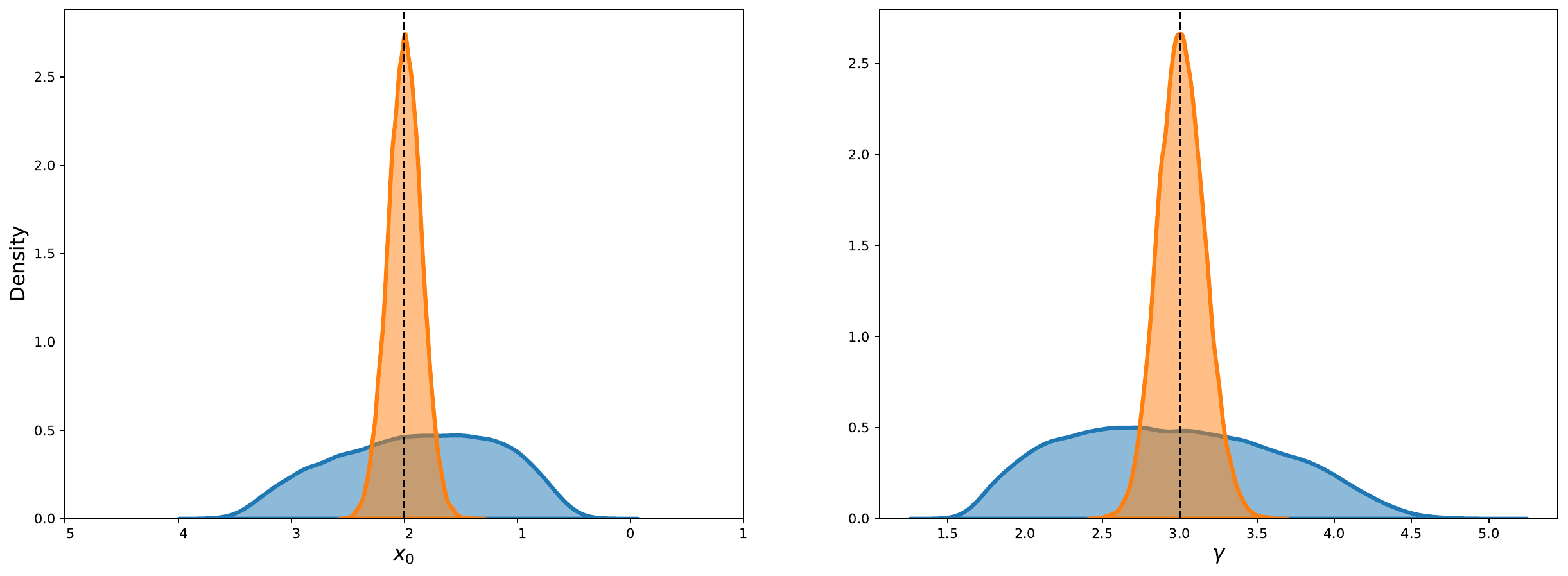}
 \caption{Posterior distributions with true Cauchy parameters $x_0 = -2$ and $\gamma = 3$, and sample size $N = 1000$, given its median and its MAD, obtained using Insufficient Gibbs Sampling (in orange) and the ABC method (in blue) with equivalent computational time, along with the theoretical values (shown as black dashed lines).}
 \label{fig:abc}
\end{figure*}

The sample median and MAD are routinely used as estimators for the location-mean Cauchy distribution, both because the Cauchy's mean and variance are undefined, and because the location parameter $x_0 \in \mathbb{R}$ is equal to the theoretical median and the scale parameter $\gamma > 0$ is equal to the theoretical MAD (and to half of the theoretical IQR). The Cauchy distribution serves as a valuable tool for exploring robust statistical methods and understanding their performance under challenging conditions, especially in the presence of heavy-tailed data or outliers. By leveraging the robust estimators of location and scale, we can overcome the limitations posed by traditional measures such as the mean and variance, and obtain more reliable estimates of the parameters of interest. 

We conducted a Metropolis-within-Gibbs random walk with Cauchy and Gamma priors on the two parameters. We performed $T$ simulations based on the posterior distribution of the Cauchy parameters $(x_0, \gamma)$ while observing only the median and the MAD.

Previous studies have resorted to Approximate Bayesian Computation (ABC) to approximate the posterior given the median and MAD \citep{turner_tutorial_2012,green_bayesian_2015, marin_relevant_2014}.
To compare our results, we also carried out simulations using standard ABC methods, using the same observed median and MAD as summary statistics, along with the same non-informative priors. We ran both algorithms for an equal amount of computing time. 
In ABC, we obtained more than 10 times more simulations by parallelizing the computations. We fixed the threshold by retaining the $T$ best simulations. We thus obtain a fair comparison, with two algorithms run for the same time leading to identical sample sizes. The results of these simulations are presented in Figure \ref{fig:abc}. Our method yields a posterior that is much more peaked around the theoretical parameter values compared to the ABC approach.
This is to be expected, since the augmented-data MCMC samples from the exact posterior, whereas ABC samples from an approximation, which typically inflates the variance.

\subsection{Bayesian model choice}

In Section \ref{sec:InsBridge}, we proposed a novel and practical bridge sampling representation based on samples from our Insufficient Gibbs Sampling method to produce
an approximation of the Bayes factor $B_{12}^\textbf{T}(\Obsrstat)$. 

Recall that this Bayes factor is only consistent for Bayesian model selection under specific conditions on the nature of the insufficient statistic. We now explore two toy examples that have been studied in the model choice literature, with consistent and inconsistent properties, respectively. The analysis of the first example highlights the importance of estimating $B_{12}^\textbf{T}(\Obsrstat)$, while the second confirms the accuracy of our estimation in practice when its theoretical value is known.
\subsubsection{Toy example 1: Normal-Laplace}
This example, introduced in \citet{relevant} to demonstrate the impact of the choice of summary statistic on the Bayes factor, compares model $\mathcal M_1$: $\Data \sim \mathcal N(\theta_1,1)$ against model $\mathcal M_2$: $\Data \sim \mathcal L(\theta_2,1/\sqrt{2})$, where the Laplace distribution has mean $\theta_2$ and scale parameter $1/\sqrt{2}$, resulting in a variance of $1$. 

Since the first and second-order moments, as well as the median, asymptotically match for these models, these summary statistics are ineffective for model selection. 
However, \citet{relevant} show that accurate model selection is possible when using the MAD as summary statistic, because its behaviour differs between the two models.
They consider synthetic data generated from each model and show that the ABC model choice posterior probabilities of each model concentrate at $0$ and $1$, respectively, as the number of observations $N$ increases.

\begin{figure*}[t!]
\centering
 \begin{subfigure}{0.3\textwidth}
  \includegraphics[width=\textwidth]{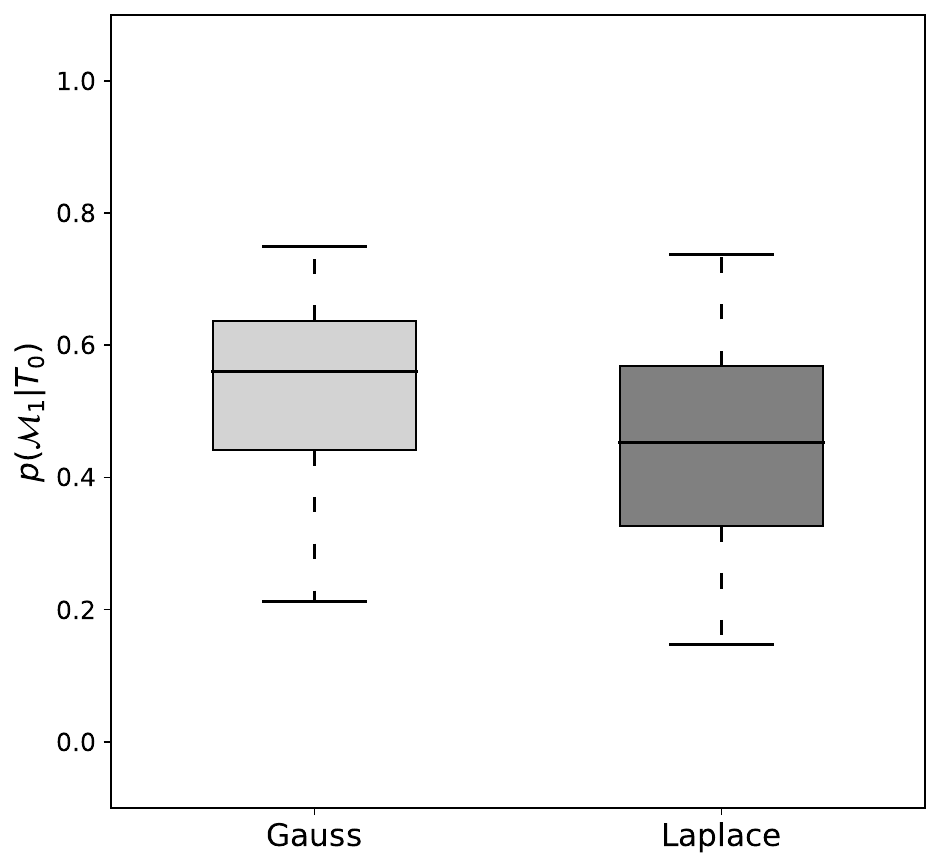}
  \caption{$N=10$}
  \label{fig:first}
 \end{subfigure}
 \hfill
 \begin{subfigure}{0.3\textwidth}
  \includegraphics[width=\textwidth]{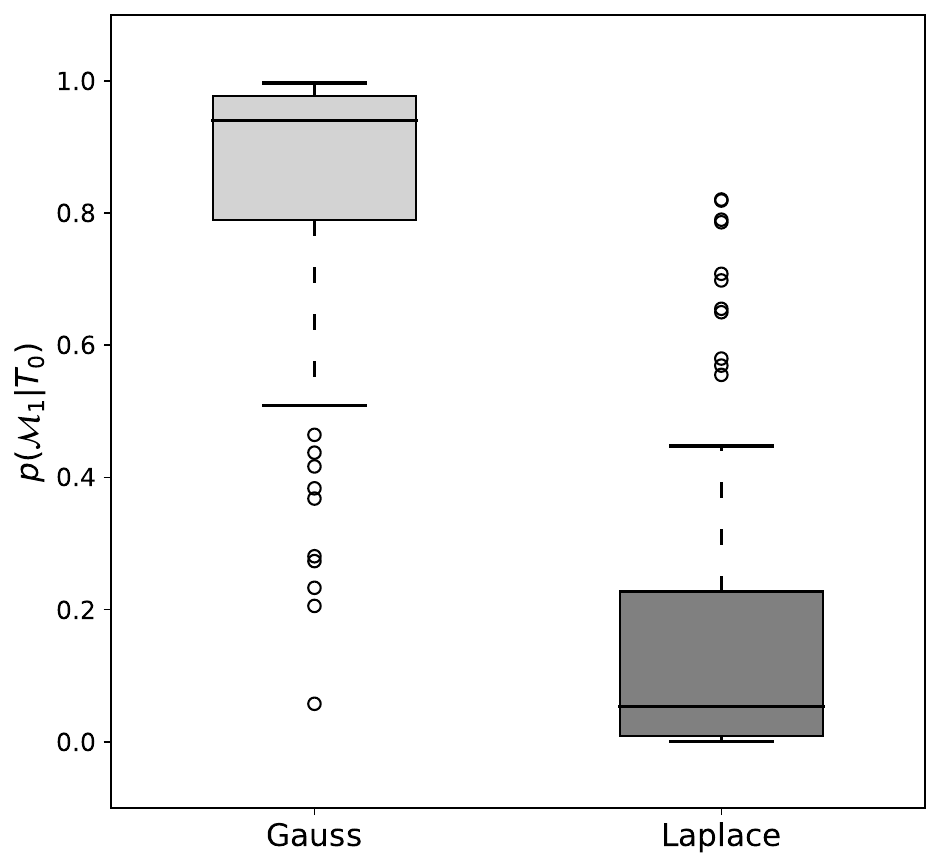}
  \caption{$N = 100$}
  \label{fig:second}
 \end{subfigure}
 \hfill
 \begin{subfigure}{0.3\textwidth}
  \includegraphics[width=\textwidth]{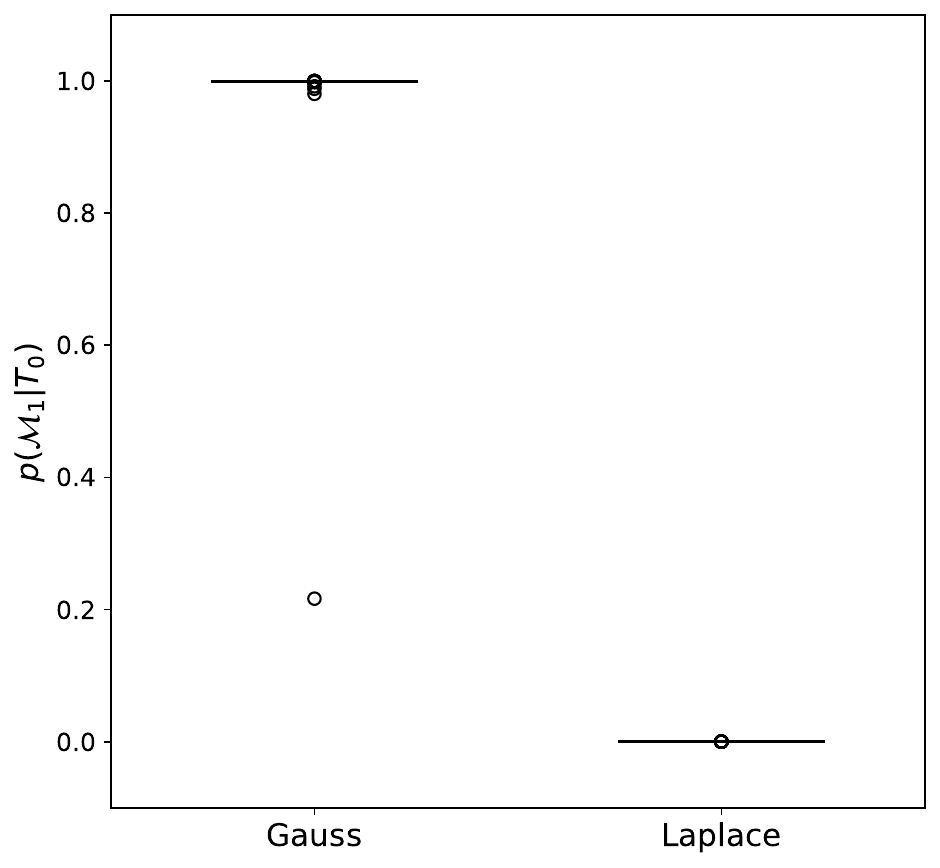}
  \caption{$N = 1000$}
  \label{fig:third}
 \end{subfigure}
 
 \caption{Comparison of the posterior probabilities supporting a Gaussian model ($\mathcal{M}_1$) with unknown mean, as the number of observations $N$ increases, from both Gaussian and Laplace simulated samples. The boxplots are derived from 100 simulation runs of samples of size $N$, and each posterior probability is obtained by insufficient bridge sampling with $T = 10^4$, conditioned on the observed statistic $\Obsrstat= \text{T}(\Data) = (\text{median}(\Data), \text{MAD}(\Data))$.}
 \label{fig:gauss_laplace}
\end{figure*}

This motivates our focus here on the case where we observe $\Obsrstat =\Rstat = (\text{median}(\Data), \text{MAD}(\Data))$, a case covered in Section \ref{sec:mad}. Here, the MAD acts as an ancillary statistic, creating a scenario where $B_{12}^\textbf{T}(\Obsrstat)$ proves consistent. Moreover, the location parameters of both models reside in the same space $\Theta$, serving identical roles and thus obviating the need for reparameterization. Hence, we denote $\theta = \theta_1 = \theta_2$. By applying our insufficient bridge sampling, we observe in Figure \ref{fig:gauss_laplace} that the distributions of the posterior probability that the
model is Normal (as opposed to Laplace) when the data are Normal and Laplace, respectively, behave as desired, concentrating near 0 and 1 as the number of observations $N$ increases. 

\subsubsection{Toy Example 2: Poisson-Geometric}

This example was introduced by \citet{grelaud} and further discussed by \citet{didelot} and \citet{Robert2011} as an illustration of the impact on Bayes factors when using a modelwise sufficient statistic. Here, we compare models $\mathcal M_1$: $\Data \sim \mathcal P(\lambda)$ and $\mathcal M_2$: $\Data \sim \mathcal G(p)$. In this scenario, the sum $S = \sum_{i=1}^N \data_i = \Rstat$ acts as a sufficient statistic within each model but not across the models. Thus, $B_{12}^\textbf{T}(\Rstat)$ takes the form \eqref{eq:suff}. We impose the conjugate priors $\lambda \sim \mathcal{E}(1)$ for the Poisson model and $p \sim \mathcal U(0,1)$ for the geometric model, which simplifies the calculation of the marginal likelihoods and thus of the Bayes factor based on this statistic:
\begin{equation*}
 B_{12}^\textbf{T}(S) = \frac{N^{S-1}(N+S)(N+S+1)}{(N+1)^{S+1}}.
\end{equation*}

Therefore, we can assess the divergence between $B_{12}^\textbf{T}(\Rstat)$ and $B_{12}(\Data)$. In this case, observing $S$ alone is insufficient for conducting convergent model choice. Nonetheless we can check the accuracy of our method against the exact value. 
To streamline the insufficient bridge sampling process, we reparameterize the geometric distribution by its mean: $\mu = \mathbb E[\Data] = \frac{1-p}{p}$. 
\begin{figure*}[t!]
 \centering
 \includegraphics[width=.8\textwidth]{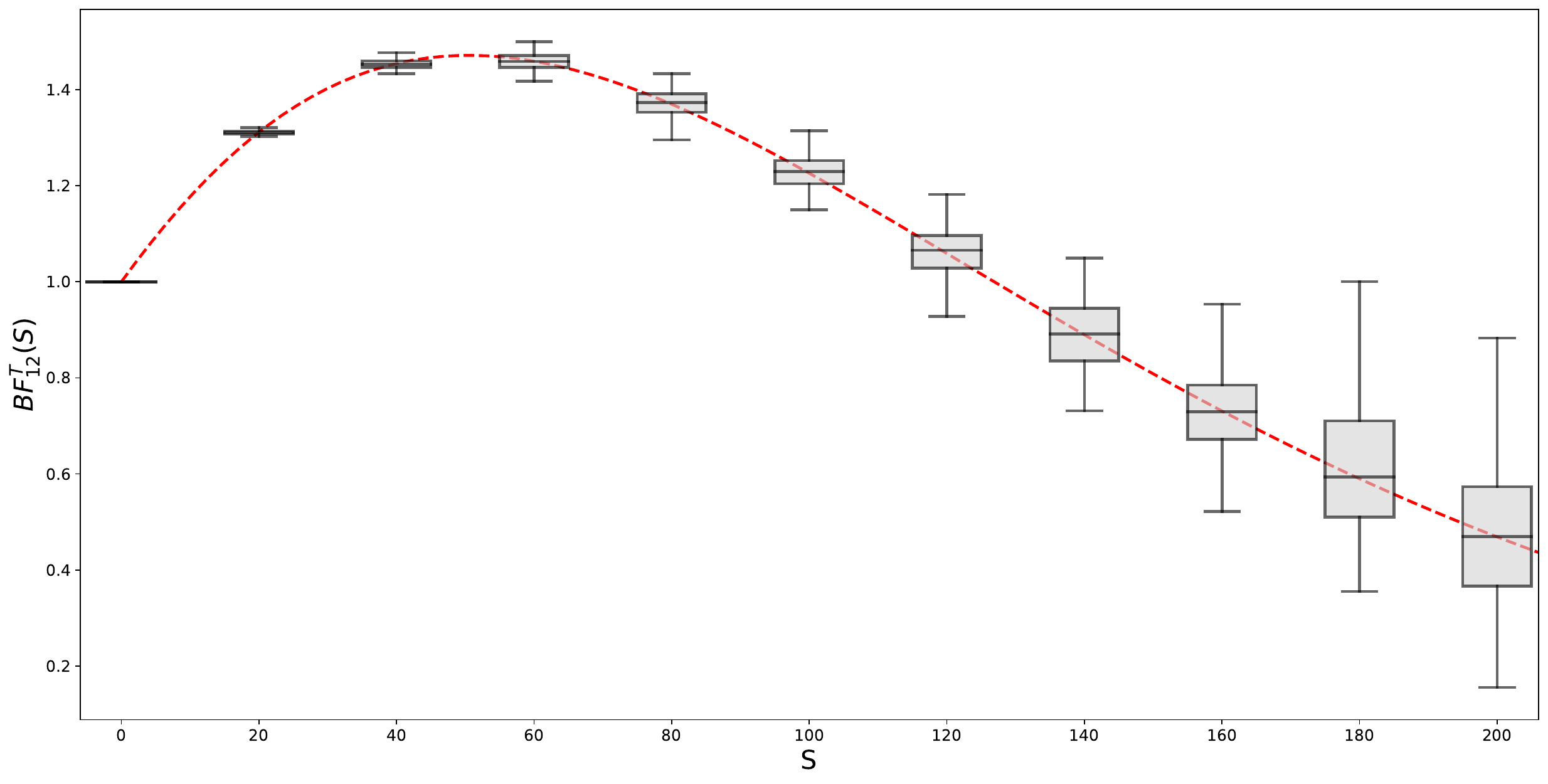}
 \caption{Distribution of the insufficient bridge sampling estimators for $B_{12}^\textbf{T}(S)$ across various values of $S$ with $\mathcal M_1$: $\Data \sim \mathcal P(\lambda)$ and $\mathcal M_2$: $\Data \sim \mathcal G(p)$ and $N = 50$. For each specified value of $S$, the Bayes Factor was estimated 100 times, utilizing $T = 10^5$ iterations.}
 \label{fig:poiss_geo}
\end{figure*}
Given $S$, the data distribution assumes a multinomial $\mathcal M(S,1/N,\dots,1/N)$ form for the Poisson model and a uniform distribution across datasets of size $N$ satisfying $\sum_i \data_i=S$ for the geometric scenario, since $S$ then follows a negative binomial $\mathcal Neg (N,p)$ distribution. Thus, straightforward simulations from both $\pi(\boldtheta \mid S)$ and $\pi(\Data \mid S)$ are available, thereby allowing for the derivation of an insufficient bridge sampling approximation. Consequently, as illustrated in Figure \ref{fig:poiss_geo}, we observe that our estimator is indeed consistent and exhibits relatively low variance for $N=50$ and various values of $S$. However, the ratio of the unnormalized posterior densities calculated for the two models incorporates the term $\left(\Pi_{i=1}^N y_i!\right)^{-1}$, which varies significantly for large $N$ and $S$. We thus caution that excessive variance in these terms may significantly increase the bias of the estimator, which can, however, be mitigated by substantially increasing the number of simulations $T$. 

% We thus caution that excessive variance in these terms may hinder the iterative scheme's ability to converge to an accurate estimator for large $N$ and $S$.

\subsection{Real data application}
\label{sec:real_data}
As outlined in the introduction, our motivation for this paper was partly driven by the inference on real-world data. To address these motivations, this section will focus on income data of the French population in 2020, publicly released by INSEE (the French official statistics agency) in the Filosofi database \citep{Filosofi2020}. The data are available at the level of the \emph{commune}, the lowest administrative division in France. For each commune, some quantiles of income data are available: the three quartiles, supplemented by the nine deciles for communes with a population over $2000$ individuals or $1000$ households. Here, we will arbitrarily focus on the commune of Contes, a small village in the Alpes-Maritimes in the South of France with $2,899$ households. Thus, we have eleven quantiles as observed statistic, i.e., $\Obsrstat = (q_j)_{j = 1,\dots,M}$ with their associated $(p_j)_{j = 1,\dots,M}$ and $M = 11$.

We first perform inference of a translated Weibull distribution of these data, as recommended by \citet{bandourian_comparison_2002}. We investigate the impact of the number of observed quantiles, denoted as $M$, on the posterior distribution. 
To achieve this, we examine subsets of the 11 quantiles with sizes $M = 2$, $M = 3$, $M = 4$, and $M = 9$, selecting the quantiles in such a way that their associated probabilities are distributed as uniformly as possible between 0 and 1 i.e $p_j \approx \frac{j}{M+1}$ for $j=1,\dots,M$. For instance, when $M=3$, we have $(p_1,p_2,p_3) = (0.25,0.5,0.75)$.

\begin{figure*}[t!]
 \centering
 \includegraphics[width=1\textwidth]{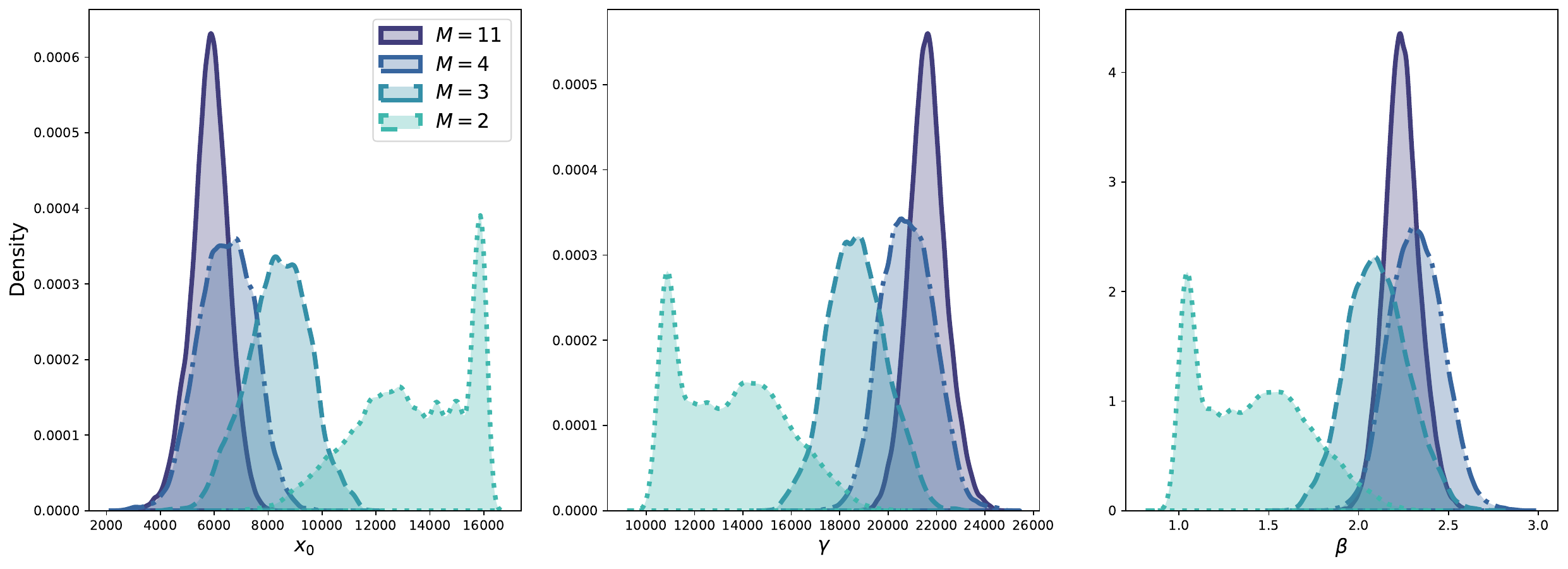}
 
 \caption{The posterior distributions of the three parameters for the translated Weibull distribution, applied to the income data from the town of Contes ($N=2899$), obtained by Insufficient Gibbs Sampling given observed quantiles $\Obsrstat=(q_j)_{j = 1,\dots,M}$ with respective probabilities $(p_j)_{j = 1,\dots,M}$ where $M = 11$. This visualization considers subsets of these quantiles with $M=2$ (depicted by a dotted line), $M=3$ (illustrated with a dashed line), $M=4$ (represented by a dash-dotted line), and the full set of $M=11$ quantiles (shown as a solid line).}
  \label{fig:real_data}
\end{figure*}

As demonstrated in Figure \ref{fig:real_data}, it is clear that relying on only two quantiles is inadequate for precisely identifying all three parameters. Yet, employing at least three quantiles allows for the accurate determination of all parameters within the Weibull distribution. Additionally, incorporating more quantiles results in a posterior distribution that is more sharply concentrated around the identified parameters, which suggests enhanced accuracy of the estimations. The analogous figure for simulated data is provided in Appendix \ref{sec:weibull}.

We next turn to model choice for these data.
In the scenario where observed quantiles are involved, an empirical approach to model selection could have involved using a QQ-plot as an exploratory tool to discern which distribution best matches the observed quantiles. With the method detailed in Section \ref{sec:BMC}, Bayesian model selection emerges as a feasible alternative or complement to such preliminary visualization through insufficient bridge sampling. It is often evident that empirical quantiles of differing distributions diverge asymptotically, thus providing a foundation for model differentiation. Hence, we proceed with a model comparison between the Lognormal and Gamma distributions, both of which are two-parameter models that can be conveniently reparameterized through their first and second moments, unlike the Weibull distribution. In this specific case, we achieve a log-Bayes Factor of $26$ in favor of the Gamma distribution. This result enables us to make a decisive choice in a coherent and justified manner.
\newpage
\section{Conclusion}

This paper has presented a novel method for simulating from the posterior distribution when only robust statistics are observed. Our approach, based on Gibbs sampling and the simulation of augmented data as latent variables, offers a versatile tool for a wide range of applied problems.
The Python code implementing this method is available as a Python package (\url{https://github.com/AntoineLuciano/Insufficient-Gibbs-Sampling}), enabling its application in various domains.

Among the three examples of robust statistics addressed in this paper, two exhibited similarities, where the observed quantiles or the median and interquartile range yielded comparable results to existing methods. The unique case of median absolute deviation (MAD) introduced a novel challenge, for which we proposed a partial data augmentation technique ensuring ergodicity of the Markov chain.
We have also leveraged the samples from both parameters and simulated data to introduce a method for conducting Bayesian model choice. Indeed, by extending the approach of the bridge sampling estimator, we can accurately estimate the Bayes Factor based solely on observed statistics, which are typically approximated by ABC model choice. Although this estimator is limited by the usual conditions of bridge sampling and the size of the simulated data, it provides a reliable and effective tool to assist statisticians in performing model choice without access to the data.

While our focus in this study was on continuous univariate distributions, future research avenues could explore the extension of our method to various observed statistics, discrete distributions, multivariate data. These directions promise to further enhance the applicability and generality of our approach.

\section*{Acknowledgements}

We are grateful to Edward I. George for a helpful discussion and in particular for suggesting the title to this paper, and to the associate editor and two anonymous reviewers for helpful feedback. Antoine Luciano is supported by a PR[AI]RIE PhD grant. Christian P. Robert is funded by the European Union under the GA 101071601, through the 2023-2029 ERC Synergy grant OCEAN and by a PR[AI]RIE chair from the Agence Nationale de la Recherche (ANR-19-P3IA-0001).

\bibliography{bibliography}
\begin{appendices}

\section{Initialization of $\Data^0 \mid \Obsrstat$}
\label{sec:init_quantile}
\subsection{Quantile case}

Our algorithm requires an initial value of the vector $\Data^0$ which verifies that $\forall j, Q(\Data^0, p_j) = q_j$.

First, we initialize the parameter vector $\boldtheta^0$ arbitrarily to enable the simulation of our vector. In order to match the observed quantiles, we set the order statistics that determine the values in the quantiles. For deterministic quantiles ($g_j=0$), we directly assign an observation equal to $q_j$. For quantiles requiring simulation, we introduce a positive distance parameter $\epsilon_j$. Specifically, for all $j$ in $J_S$, we set $\data_{\scriptscriptstyle{(i_j)}}=q_j-\epsilon_j g_j$ and $\data_{\scriptscriptstyle{(i_j+1)}}=q_j+\epsilon_j (1-g_j)$, ensuring that $g_j \data_{\scriptscriptstyle{(i_j+1)}}+(1-g_j)\data_{\scriptscriptstyle{(i_j)}}=q_j$. To enhance the efficiency of our initialization, we can normalize $\epsilon_j$ to be equal to the variance of $\data_{\scriptscriptstyle{(i_j)}}$ under the assumption that $\Data$ follows a distribution denoted as $\mathcal{F}_{\boldtheta^0}$. Once these observations have been initialized, we can complete the initial vector $\Data^0$ by simulating the remaining observations in the appropriate intervals using a truncated distribution $\mathcal{F}_{\boldtheta^0}$.

\subsection{Median and IQR case}
\label{sec:init_iqr}

We must initialize our MCMC with values $(\Data^0, \boldtheta^0)$ that verify the constraints $\text{median}(\Data^0) = m, \text{IQR}(\Data^0) = i$, and $\prod_j f_{\boldtheta^0}(\Data^0_j)>0$. 

If the family $(\mathcal F_{\boldtheta})_{\boldtheta \in \boldtheta}$ has support the whole real line, we simulate a vector $\textbfz$ of size $N$ from an arbitrary distribution (such as $\mathcal N(0,1)$ or $\mathcal F_{\boldtheta^0}$) and then apply a linear transformation to it so that it verifies the constraints: 
$\text{median(}\Data^0\text{)}=m\text{ and IQR(}\Data^0\text{)}=i$. So we have:

$$\Data^0 = (\textbfz- \text{median}(\textbfz)) \frac{i}{\text{IQR}(\textbfz)}+m.$$

In the case where the distribution is defined on a strict subset of $\mathbb{R}$, this technique is inappropriate, as it may lead to initial values which lie outside the support of the distribution. 
In this situation, we use a deterministic initialization instead. 

The initialization vector is then given by:

\begin{eqnarray*}
\data_1=\data_2=\ldots=\data_{ n} &=& m-\frac{3i}{4}\\
\data_{n+1} &=& q_1\\
\data_{n+2}=\ldots=\data_{2n} &=& m-\frac i4\\
\data_{2n+1} &=& m\\
\data_{2n+2}=\ldots=\data_{3n } &=& m+\frac i4\\
\data_{3n+1} &=&q_3\\
\data_{3n+2} = \ldots = \data_{4n+1} &=& m+\frac{3i}{4},
\end{eqnarray*}
assuming that all these values have positive density under $f_{\boldtheta^0}$.
This initial vector verifies the constraints. In practice, with this initialization, we find that a burn-in time of about $5N$ iterations is sufficient to reach the stationary distribution.

\subsection{Median and MAD case}
\label{sec:init_mad}

As with the case where we observe the median and the IQR presented in in the previous section, we introduce two techniques for initializing the vector $\Data^0$. The first, and default, technique consists in simulating a vector $\textbfz$ of size $N$ from an arbitrary distribution (such as $\mathcal N(0,1)$ or $\mathcal F_{\boldtheta^0}$) and then applying a linear transformation to it so that it verifies the constraints: 
$\text{median(}\Data^0\text{)}=m\text{ and MAD(}\Data^0\text{)}=s$. So we have:

$$\Data^0 = (\textbfz- \text{median}(\textbfz)) \frac{s}{\text{MAD}(\textbfz)}+m.$$ 

In our numerical experiments following this method, we observe that the burn-in period is extremely short.

As in the IQR scenario, if the support of the distribution is a strict subset of $\mathbb{R}$, we resort to the deterministic initialization, which corresponds to an apportionment with $k = \lceil \frac{n}{2} \rceil= \lceil \frac{N-1}{4} \rceil$ and $\delta = 1$. Therefore, we define:

\begin{eqnarray*}
\data_1=\data_2=\ldots=\data_{ n-k+1} &=& m-\frac{3s}{2}\\
\data_{n-k+2}=\ldots=\data_n &=& m-\frac s2\\
\data_{n+1} &=& m\\
\data_{n+2}=\ldots=\data_{ 2n-k+1 } &=& m+\frac s2\\
\data_{2n-k+2} &=& m+s\\
\data_{2n-k+3} = \ldots = \data_{2n+1} &=& m+\frac{3s}{2},
\end{eqnarray*}
assuming that these values all lie within the support of the distribution.
This initial vector verifies the constraints. In practice, with this initialization, we find that a burn-in time of about $5N$ iterations is sufficient to reach stationarity.

\section{Other median, IQR cases}
In Section \ref{sec:iqr}, we outlined a method for simulating \(\Data\) that meets the criteria \(\text{median}(\Data) = m \in \mathbb R\) and \(\text{IQR}(\Data) = i>0\) when \(N = 4n+1\). The method remains substantially the same; here, we describe the models associated with variable transformations in the three other possible cases.

\label{sec:appendi\data_iqr}
\subsection{$N = 3 \mod 4$}

In this scenario, when $N = 3n+4$ with $n\in \mathbb{N}^*$, the median is deterministic, and we have $ \text{median}(\Data) = \data{\scriptscriptstyle(i_2)}$. But this time, the two other quartiles are not deterministic. We have $Q_1 = (1 - g_1)\data{\scriptscriptstyle(i_1)} + g_1 \data{\scriptscriptstyle(i_1+1)}$ and $Q_3 = (1 - g_3)\data{\scriptscriptstyle(i_3)} + g_3 \data{\scriptscriptstyle(i_3+1)}$, with $i_1 = k+1$, $i_2 = 2k+2$, $i_3 = 3k+2$, and $g_1 = g_3 = 0.5$. We have to consider the joint distribution of five order statistics $\data{\scriptscriptstyle(i_1)}, \data{\scriptscriptstyle(i_1+1)}, \data{\scriptscriptstyle(i_2)}, \data{\scriptscriptstyle(i_3)}, \data{\scriptscriptstyle(i_3+1)}$ and apply the transformation described in the second system:

$$\left\{
 \begin{aligned} 
 Q_1 &= (1 - g_1)\data{\scriptscriptstyle(i_1)} + g_1 \data{\scriptscriptstyle(i_1+1)} \\
 m &= \data{\scriptscriptstyle(i_2)} \\
 Q_3 &= (1 - g_3)\data{\scriptscriptstyle(i_3)} + g_3 \data{\scriptscriptstyle(i_3+1)} \\
 i &= Q_3 - Q_1
 \end{aligned}
 \right. \\
\iff  \left\{
 \begin{aligned}
 \data{\scriptscriptstyle(i_1)} &= \data{\scriptscriptstyle(i_1)} \\
 \data{\scriptscriptstyle(i_1+1)} &= \frac{1}{g_1} \large( (1 - g_3)\data{\scriptscriptstyle(i_3)} + g_3 \data{\scriptscriptstyle(i_3+1)}\\
 & \quad - (1 - g_1) \data{\scriptscriptstyle(i_1)} - i \large) \\
 \data{\scriptscriptstyle(i_2)} &= m \\
 \data{\scriptscriptstyle(i_3)} &= \data{\scriptscriptstyle(i_3)} \\
 \data{\scriptscriptstyle(i_3+1)} &= \data{\scriptscriptstyle(i_3+1)}
 \end{aligned}
 \right.
$$

% In the same way as before, we can sample from the distribution of $\data{\scriptscriptstyle(i_1)}, \data{\scriptscriptstyle(i_3)}, \data_{((i_3+1))} \mid \text{median}(\Data) = m, \text{IQR}(\Data) = i$ with a step of Metropolis-Hastings.

\subsection{$N$ is even}

In this case, which actually covers two different cases (when $N = 4k$ or $N = 4k+2$), none of the three quartiles is deterministic. The values of the median and the IQR then depend on six order statistics $\data{\scriptscriptstyle(i_1)}, \data{\scriptscriptstyle(i_1+1)}, \data{\scriptscriptstyle(i_2)}, \data{\scriptscriptstyle(i_2+1)}, \data{\scriptscriptstyle(i_3)}, \data{\scriptscriptstyle(i_3+1)}$. We have:
$$\left\{
 \begin{aligned} 
 Q_1 &= (1 - g_1)\data{\scriptscriptstyle(i_1)} + g_1 \data{\scriptscriptstyle(i_1+1)} \\
 m &= \frac{\data{\scriptscriptstyle(i_2)} + \data{\scriptscriptstyle(i_2+1)}}{2} \\
 Q_3 &= (1 - g_3)\data{\scriptscriptstyle(i_3)} + g_3 \data{\scriptscriptstyle(i_3+1)} \\
 i &= Q_3 - Q_1
 \end{aligned}
 \right. \\
 \iff \left\{
 \begin{aligned}
 \data{\scriptscriptstyle(i_1)} &= \data{\scriptscriptstyle(i_1)} \\
 \data{\scriptscriptstyle(i_1+1)} &= \frac{1}{g_1} ( (1 - g_3)\data{\scriptscriptstyle(i_3)} + g_3 \data{\scriptscriptstyle(i_3+1)}\\
 & \quad - (1 - g_1) \data{\scriptscriptstyle(i_1)} - i) \\
 \data{\scriptscriptstyle(i_2)} &= \data{\scriptscriptstyle(i_2)} \\
 \data{\scriptscriptstyle(i_2+1)} &= 2m - \data{\scriptscriptstyle(i_2)} \\
 \data{\scriptscriptstyle(i_3)} &= \data{\scriptscriptstyle(i_3)} \\
 \data{\scriptscriptstyle(i_3+1)} &= \data{\scriptscriptstyle(i_3+1)}
 \end{aligned}
 \right.
$$

When $N = 4k$, we have $i_1 = k$, $i_2 = 2k$, $i_3 = 3k$, $g_1 = 0.75$, and $g_3 = 0.25$. When $N = 4k+2$, we have $i_1 = k+1$, $i_2 = 2k+1$, $i_3 = 3k+1$, $g_1 = 0.25$, and $g_3 = 0.75$.

\section{Proof of ergodicity for the chain of latent vector $\Data$}
\label{sec:ergodic}

In this section, we prove that our Markov chain on the latent vector $\Data$ is ergodic. To achieve this, we need to demonstrate its irreducibility and aperiodicity.

Aperiodicity is ensured by the random selection of the coordinates to be resampled and the inherent randomness in the simulation process.

To establish irreducibility, we need to show that all states are reachable within a finite number of iterations. In this regard, we have defined two variables, $k$ and $\delta$, which describe the distribution of observations in the $\Data$ vector. Therefore, it suffices to demonstrate that we can transition from one state to any other state from this latent space $\mathcal{x}_{m,s}=\{x \in \mathbb{R}^N \mid \text{median}(\Data)=m, \text{MAD}(\Data)=s\}$.

Let us consider two states of our Markov chain on the latent vector from $\mathcal{x}_{m,s}$, $\Data$ such that $k(\Data) = k_1 \in \{1,\dots,n\}$ and $\delta(\Data) = \delta_1 \in \{0, 1\}$ to any other state $\Tilde{\Data}$ such that $k(\Tilde x) = k_2 \in \{1,\dots,n\}$ and $\delta(\Tilde \Data) = \delta_2 \in \{0, 1\}$.

If $k_1 \neq k_2$, we can assume with no loss of generality that $k_2 > k_1$. In that case, we need to take $k_2-k_1$ observations in $(-\infty,m-s)$ and $k_2-k_1$ in $(m,m+s)$ and resample them respectively in $(m-s,m)$ and $(m+s,+\infty)$. This case has a non-zero probability to occur and is described as case 4a in our algorithm. 

If $k_1 = k_2$ and $\delta_1 \neq \delta_2$, we need to switch $\data_{\scriptscriptstyle\mathrm{MAD}}$ from one side of the median to its symmetric side, i.e., $m-s \leftrightarrow m+s$. This case has a non-zero probability to occur and is described as case 3b in our method.

Finally, if $k_1 = k_2$ and $\delta_1 = \delta_2$, $\Data$ and $\Tilde \Data$ present the same distribution between the four intervals introduced previously. We have to resample the observations while maintaining this distribution. This case has a non-zero probability to occur and is described as cases 4b, 3a, and 2 in our method.

By considering these possibilities, we can establish the existence of a non-zero probability for transitioning between any two states in the specified latent space.

Therefore, we have demonstrated the irreducibility of our Markov chain, indicating that it can reach all states within a finite number of iterations.
\section{Even case of the median, MAD case}
\label{sec:even}

As mentioned in Section \ref{sec:mad}, the case where $\Data$ is of even size i.e. $\exists n\geq 4 \mbox{ such that }N=2n$ is more complex. Indeed, the statistics that we study are not determined by a single coordinate but by the average of two coordinates of the vector. 

It is therefore useful to introduce some extra notations: 

\begin{itemize}
 \item[-] $m_1 = \data_{(n)}$ and $m_2=\data_{(n+1)}$ where $\data_{(k)}$ denotes the k-th order statistic of the vector $\Data$, so we have $m=\frac{m_1+m_2}{2}$.
 % \item[-] $Y= (\lvert \data_i-m\rvert)_{i=1,\dots,n}$ the vector of distance to the median 
 \item[-] $s_1=\data_{(n)}$ and $s_2=\data_{(n+1)}$ so we have $s=\frac{s_1+s_2}{2}$
 \item[-] $\data_{\scriptscriptstyle{\text{MAD,1}}}\in \Data$ such that $\lvert \data_{{\scriptscriptstyle{\text{MAD,1}}}} - m \rvert = s_1$ and $\data_{\scriptscriptstyle{\text{MAD,2}}}\in \Data$ such that $\lvert \data_{\scriptscriptstyle{\text{MAD,2}}} - m \rvert = s_2$
 \item[-] $\epsilon_m = m_2-m_1$ and $\epsilon_{\scriptscriptstyle{\text{MAD}}}=s_2-s_1$
\end{itemize}

\begin{figure*}[!h]
\begin{center} 
 \begin{tikzpicture}[scale=1.8]
 \begin{axis}[%
 axis x line=center,
 axis y line=none,
 xmin=0,xmax=1, ymin=0, ymax=3,
 xtick={0.2,0.25,0.3,.45,0.5,.55,.7,0.75,.8},
 xticklabels={},
 ]
 \end{axis}
 \node [draw=none] at (.75,.35) {$Z_1$} ;
 \node [draw=none] at (1.75,0.35) {$Z_{m-s}$} ;
 \node [draw=none] at (2.55,0.35) {$Z_2$} ;
 \node [draw=none] at (3.4,0.35) {$Z_m$} ;
 \node [draw=none] at (4.3,0.35) {$Z_3$} ;
 \node [draw=none] at (5.15,0.35) {$Z_{m+s}$} ;
 \node [draw=none] at (6,0.35) {$Z_4$} ;

 \node [draw=none] at (1.37,-.4) {$m-s_2$} ;
 \node [draw=none] at (1.68,-0.2) {$m-s$} ;
 \node [draw=none] at (2.07,-0.4) {$m-s_1$} ;
 \node [draw=none] at (3.04,-0.4) {$m_1$} ;
 \node [draw=none] at (3.42,-0.2) {$m$} ;
 \node [draw=none] at (3.8,-0.4) {$m_2$} ;
 \node [draw=none] at (4.79,-.4) {$m+s_1$} ;
 \node [draw=none] at (5.1,-0.2) {$m+s$} ;
 \node [draw=none] at (5.48,-0.4) {$m+s_2$} ;

 \end{tikzpicture}
 
 \end{center}
 \caption{Apportionment of the vector $\Data$ when $N=2n$ with $n\in \mathbb{N}^*$, median$(\Data)=m$ and MAD$(\Data)=s$.}
 \label{fig:mad_even}
\end{figure*}
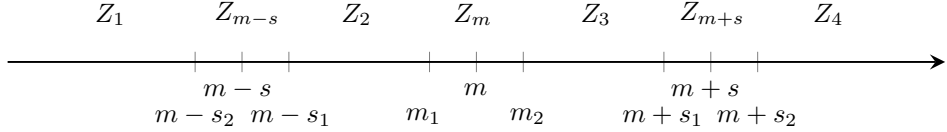

As for the odd case, we introduce a new partition of $\mathbb{R}$ but this time involving seven intervals presented in \ref{fig:mad_even}. 
The intervals are denoted as follows: 

\begin{itemize}
 \item[-] $Z_1 = (-\infty,m-s_2)$
 \item[-] $Z_{m-s} = (m-s_2,m-s_1)$
 \item[-] $Z_2 = (m-s_1,m_1)$
 \item[-] $Z_{m} = (m_1,m_2)$
 \item[-] $Z_3 = (m_2,m+s_1)$
 \item[-] $Z_{m+s} = (m+s_1,m+s_2)$
 \item[-] $Z_4 = (m+s_2,+\infty)$
\end{itemize}

We use the notations from the previous section applied to our new zones. In addition, we introduce:

\begin{itemize}
 \item[-] An application \textbf{$S_x$} which associates to a point $\data$ its symmetric with respect to the point $x$ i.e $S_x: x \longrightarrow 2x-\data$.
 \item[-] An application Zone which has a coordinate associates its zone.
 \item[-] An application $\textrm{Zone}_E$ which has a coordinate associates its ``extended" zone to it such that
 
 \vspace{.1cm}
 
 $\textrm{Zone}_E(\data) = \left\{
 \begin{array}{ll}
  (-\infty,m-s) & \mbox{if } \data\in Z_1 \\
  (m-s,m_1) & \mbox{if } \data\in Z_2 \\
  (m_2,m+s) & \mbox{if } \data\in Z_3 \\
  (m+s,\infty) & \mbox{if } \data\in Z_4 \\
 \end{array}
\right.
$
 \item[-] An application $\textrm{Zone}_S$ which has a coordinate associates its ``symmetric" zone with respect to $m$ such that $\textrm{Zone}_S(\data) = \left\{
 \begin{array}{ll}
  Z_4 & \mbox{if } \data\in Z_1 \\
  Z_3 & \mbox{if } \data\in Z_2 \\
  Z_2 & \mbox{if } \data\in Z_3 \\
  Z_1 & \mbox{if } \data\in Z_4 \\
 \end{array}
\right.
$
 \item[-] An application $\textrm{Zone}_C$ which has a zone associates its ``complementary" zone to it such that $\textrm{Zone}_C(\data) = \left\{
 \begin{array}{ll}
  Z_3 & \mbox{if } \data\in Z_1 \\
  Z_4 & \mbox{if } \data\in Z_2 \\
  Z_1 & \mbox{if } \data\in Z_3 \\
  Z_2 & \mbox{if } \data\in Z_4 \\
 \end{array}
\right.
$
\end{itemize}

The algorithm for sampling \((\tilde \data_i, \tilde \data_j)\) from the joint distribution, after randomly selecting \((\data_i, \data_j)\), is as follows:
\begin{enumerate}

 \item If $(\data_i, \data_j) = (m_1,m_2)$: 
 \begin{itemize}
  \item[-] $\Tilde{\data_i} \sim \mathcal{F}_{\boldtheta} \mathbb{1}_{[m-\data_{(3)},m+\data_{(3)}]}$
  \item[-] $\Tilde{\data_j}= S_m(\Tilde{\data_i})$
 \end{itemize}
 \item If $(\data_i, \data_j) = (\data_{\text{MAD}_1},\data_{\scriptstyle{\text{MAD}}_2})$: 
 \begin{enumerate}
  \item If $\data_i>m \text{ and }\data_j>m$: 
  \begin{itemize}
   \item[-] $\Tilde{\data_i}\sim \mathcal{F}_{\boldtheta}\mathbb{1}_{Z_{m+s}'} \text{ where } Z_{m+s}' = [m+s_1-\epsilon,m+s_2+\epsilon]\newline \text{and } \epsilon=\min(\data_{(n)}-\data_{(n-1)},\data_{(n+2)}-\data_{(n+1)})$
   \item[-] $\Tilde{\data_j} = S_{m+s}(\Tilde{\data_i})$
  
  \end{itemize}
  \item Elif $\data_i<m \text{ and }\data_j<m$:
  \begin{itemize}
   \item[-] $\Tilde{\data_i}\sim \mathcal{F}_{\boldtheta}\mathbb{1}_{Z_{m-s}'} \text{ where } Z_{m-s}' = [m-s_2-\epsilon,m-s_1+\epsilon]\newline \text{and } \epsilon=\min(\data_{(n)}-\data_{(n-1)},\data_{(n+2)}-\data_{(n+1)})$
   \item[-] $\Tilde{\data_j} = S_{m-s}(\Tilde{\data_i})$
  \end{itemize}  
   \item Else 
   \begin{itemize}
   \item[-] $\Tilde{\data_i}\sim \mathcal{F}_{\boldtheta}\mathbb{1}_{Z_{m-s}'\cup Z_{m+s}'}$
   \item[-] $\Tilde{\data_j} = \left\{
 \begin{array}{ll}
  S_{m+s}(S_m(\Tilde{\data_i})) & \mbox{if } \Tilde{\data_i}>m \\
  S_{m-s}(S_m(\Tilde{\data_i})) & \mbox{else}
 \end{array}
\right.
$
   \end{itemize}
 \end{enumerate}
 \item If $(\data_i,\data_j) \in \{m_1,m_2\}\times \{\data_{\scriptscriptstyle{\text{MAD,1}}},\data_{\scriptscriptstyle{\text{MAD,2}}}\}$:
 \item If $\data_i = m_1\text{ or } m_2$:
 \begin{enumerate}
  \item If $\data_i=m_1$ and $\data_j \in [m_2 , m-s_1] $:
  \begin{itemize}
   \item[-] $\Tilde{\data_i} \sim \mathcal{F}_{\boldtheta} \mathbb{1}_{[m,m-s_1]}$
   \item[-] $\Tilde{\data_j} = \left\{
 \begin{array}{ll}
  S_{m}(\Tilde{\data_i}) & \mbox{if } \Tilde{\data_i}\in [m_1,m_2] \\
  \data_j & \mbox{else}
 \end{array}
\right.
$
\end{itemize}
\item Elif $\data_i=m_2$ and $\data_j \in [m-s_1 , m_1] $:
  \begin{itemize}
   \item[-] $\Tilde{\data_i} \sim \mathcal{F}_{\boldtheta} \mathbb{1}_{[m-s_1,m]}$
   \item[-] $\Tilde{\data_j} = \left\{
 \begin{array}{ll}
  S_{m}(\Tilde{\data_i}) & \mbox{if } \Tilde{\data_i}\in [m_1,m_2] \\
  \data_j & \mbox{else}
 \end{array}
\right.
$
\end{itemize}
  \item Else: $\Tilde{\data_i} \sim \mathcal{F}_{\boldtheta} \mathbb{1}_{\textrm{Zone}(\data_i)} \text{ and }\Tilde{\data_j}=\data_j$:

 \end{enumerate}
 
 \item If $\data_j = \data_{\scriptscriptstyle{\text{MAD,1}}} \text{ or }\data_{\scriptscriptstyle{\text{MAD,2}}}$:
 \begin{enumerate}
  \item If $(\data_i-m)(\data_j-m)>0$ and $(\lvert \data_i-m\rvert - s)(\lvert \data_j-m\rvert - s)<0$:
  \begin{itemize}
   \item[-] $\data_i \sim \mathcal{F}_{\boldtheta} \mathbb{1}_{\textrm{Zone}_E(\data_i)}$ 
   \item[-] $\Tilde{\data_j} = \left\{
 \begin{array}{ll}
  S_{m+s}(\Tilde{\data_i}) & \mbox{if } \Tilde{\data_i}\in Z_{m+s} \\
  S_{m-s}(\Tilde{\data_i}) & \mbox{if } \Tilde{\data_i}\in Z_{m-s} \\
  \data_j & \mbox{else}
 \end{array}
\right.
$
  \end{itemize}
  \item Elif $(\data_i-m)(\data_j-m)>0$ and $(\lvert \data_i-m\rvert - s)(\lvert \data_j-m\rvert - s)>0$:
  \begin{itemize}
   \item[-] $\Tilde{\data_i} \sim \mathcal{F}_{\boldtheta} \mathbb{1}_{\textrm{Zone}(\data_i)}$
   \item[-] $\Tilde{\data_j}=\data_j$
  \end{itemize}
  \item Elif $(\data_i-m)(\data_j-m)<0$ and $(\lvert \data_i-m\rvert - s)(\lvert \data_j-m\rvert - s)>0$:
  \begin{itemize}
   \item[-] $\Tilde{\data_i} \sim \mathcal{F}_{\boldtheta} \mathbb{1}_{\textrm{Zone}(\data_i) \cup \textrm{Zone}_S(\data_i)}$
   \item[-] $\Tilde{\data_j} = \left\{
 \begin{array}{ll}
  \data_j & \mbox{if } \Tilde{\data_i}\in \textrm{Zone}(\data_i) \\
  S_m(\data_j) & \mbox{if }\Tilde{\data_i}\in \textrm{Zone}_S(\data_i)
 \end{array}
\right.
$
  \end{itemize}
  \item Else: \begin{itemize}
   \item[-] $\Tilde{\data_i} \sim \mathcal{F}_{\boldtheta} \mathbb{1}_{\textrm{Zone}_E(\data_i)}$
   \item[-] $\Tilde{\data_j} = \left\{
 \begin{array}{ll}
  S_{m+s}(S_m(\Tilde{\data_i})) & \mbox{if } \Tilde{\data_i}\in Z_{m-s} \\
  S_{m-s}(S_m(\Tilde{\data_i})) & \mbox{if } \Tilde{\data_i}\in Z_{m+s} \\
  \data_j & \mbox{else}
 \end{array}
\right.
$
  \end{itemize}
 \end{enumerate}
 
 \item Else:
 \begin{enumerate}
  \item If $(\textrm{Zone}(\data_i),\textrm{Zone}(\data_j))=(Z_1,Z_2)$ or $(Z_3,Z_4)$:
  \begin{itemize}
   \item[-] $\Tilde{\data_i} \sim \mathcal{F}_{\boldtheta}\mathbb{1}_{\textrm{Zone}_E(\data_i)\cup \textrm{Zone}_E(\data_j)}$
   \item[-] $\left\{
 \begin{array}{ll}
  \Tilde{\data_j}=S_{m+s}(\Tilde{\data_i}) & \mbox{if } \Tilde{\data_i}\in Z_{m+s} \\
  \Tilde{\data_j}=S_{m-s}(\Tilde{\data_i}) & \mbox{if } \Tilde{\data_i}\in Z_{m-s} \\
  \Tilde{\data_j} \sim \mathcal{F}_{\boldtheta}\mathbb{1}_{\textrm{Zone}(\data_j)} & \mbox{if } \Tilde{\data_i}\in \textrm{Zone}(\data_i) \\
  \Tilde{\data_j} \sim \mathcal{F}_{\boldtheta}\mathbb{1}_{\textrm{Zone}(\data_i)} & \mbox{else } 
 \end{array}
\right.
$
  \end{itemize}
  \item Elif $(\textrm{Zone}(\data_i),\textrm{Zone}(\data_j))=(Z_2,Z_3)$:
  \begin{itemize}
   \item $\Tilde{\data_i} \sim \mathcal{F}_{\boldtheta}\mathbb{1}_{Z_2\cup Z_m \cup Z_3}$
   \item $\left\{
 \begin{array}{ll}
  \Tilde{\data_j}=S_{m}(\Tilde{\data_i}) & \mbox{if } \Tilde{\data_i}\in Z_{m} \\
  \Tilde{\data_j} \sim \mathcal{F}_{\boldtheta}\mathbb{1}_{\textrm{Zone}(\data_j)} & \mbox{if } \Tilde{\data_i}\in \textrm{Zone}(\data_i) \\
  \Tilde{\data_j} \sim \mathcal{F}_{\boldtheta}\mathbb{1}_{\textrm{Zone}(\data_i)} & \mbox{else } 
 \end{array}
\right.
$
  \end{itemize}
  \item Elif $(\textrm{Zone}(\data_i),\textrm{Zone}(\data_j))=(Z_1,Z_3)$ or $(Z_2,Z_4)$:
  \begin{itemize}
   \item[-] $\Tilde{\data_i} \sim \mathcal{F}_{\boldtheta}\mathbb{1}_{\mathbb{R}\setminus Z_m}$
   \item[-] $\left\{
 \begin{array}{ll}
  \Tilde{\data_j}=S_{m+s}(S_m(\Tilde{\data_i})) & \mbox{if } \Tilde{\data_i}\in Z_{m-s} \\
  \Tilde{\data_j}=S_{m-s}(S_m(\Tilde{\data_i})) & \mbox{if } \Tilde{\data_i}\in Z_{m+s} \\
  \Tilde{\data_j} \sim \mathcal{F}_{\boldtheta}\mathbb{1}_{\textrm{Zone}_C(\Tilde{\data_i})} & \mbox{else } 
 \end{array}
\right.
$
  \end{itemize}
  
  \item Else: $\Tilde{\data_i} \sim \mathcal{F}_{\boldtheta} \mathbb{1}_{\textrm{Zone}(\data_i)}$ and $\Tilde{\data_j}=\data_j$.
 \end{enumerate}
\end{enumerate}
$(\data_i,\data_j) = (\Tilde{\data_i},\Tilde{\data_j})$

\section{Example: Weibull distribution} 
\label{sec:weibull}
\begin{figure*}[!h]
 \centering
 \includegraphics[width=1\textwidth]{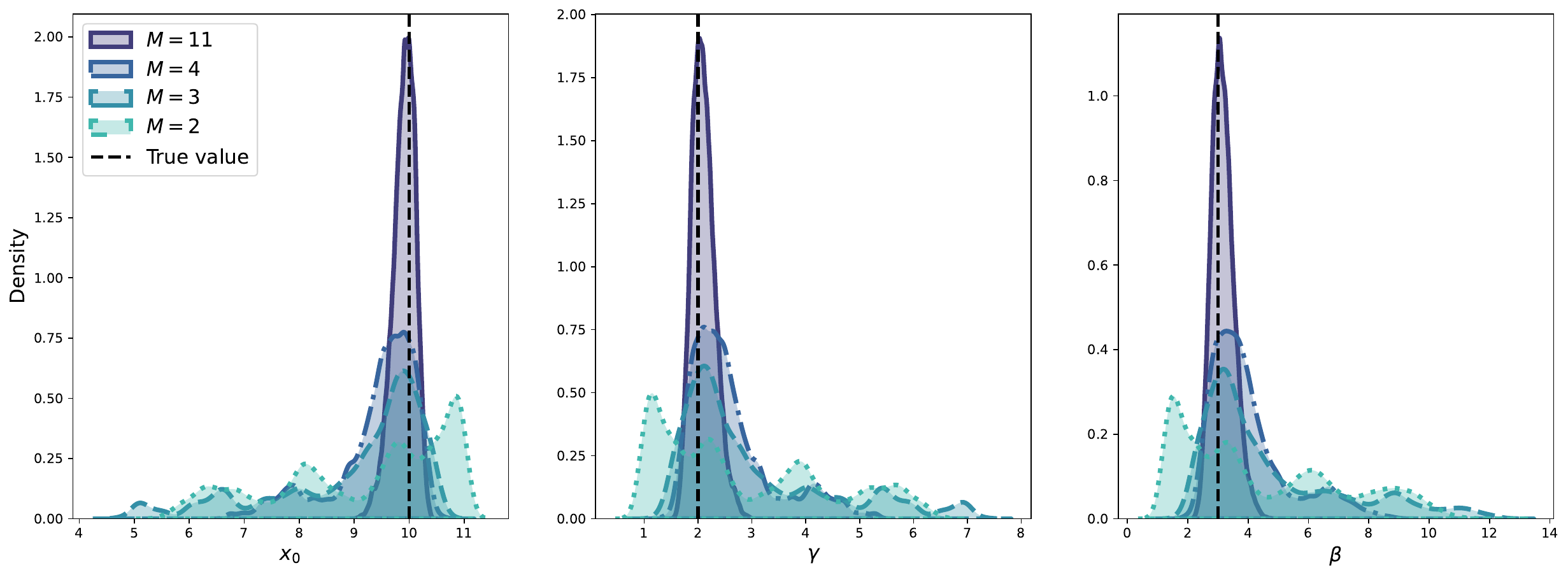}
 
 \caption{Posterior distribution of the three parameters of the Weibull distribution for a sample size of $N=1000$, obtained via Insufficient Gibbs Sampling using $\Rstat=((q_j)_{j = 1,\dots,M},(p_j)_{j = 1,\dots,M})$, where the $(q_j)$s represent the theoretical quantiles of the 3-parameter Weibull distribution with $x_0=10$, $\gamma =2$, and $\beta=3$ (depicted by black dashed lines) for various numbers of observed quantiles: $M=2$ (depicted by a dotted line), $M=3$ (illustrated with a dashed line), $M=4$ (represented by a dash-dotted line), and $M=11$ (shown as a solid line).}

 \label{fig:weibull_theoretical}
\end{figure*}
In this section, we focus on the three-parameter Weibull distribution, also referred to as the translated Weibull distribution. In addition to the classical parameters of scale $\gamma$ and shape $\beta$, this one proposes a parameter of location $x_0$. The density of this family of distributions is given by: 
$$f(x) = \frac{\beta}{\gamma} \left(\frac{x-x_0}{\gamma}\right)^{\beta-1} e^{-(\frac{x-x_0}{\gamma})^\beta} \mathbb{1}_{x\geq x_0}.$$
\citet{bandourian_comparison_2002} recommend this distribution to model the life expectancy or income of individuals.

When we observe a two-dimensional summary statistic $\Rstat$, such as the median and either the MAD or the IQR, the information contained in $\Rstat$ does not allow us to identify all three parameters. The median provides information about the location parameter, while the MAD or IQR gives information about the scale parameter. However, there is no direct information about the shape parameter. Therefore, we have an insufficient number of statistics to estimate all three parameters accurately. 
%This leads to a multitude of possible parameter combinations that can produce the same median and MAD. 
As a result, our three chains can evolve within a submanifold of $\mathbb{R}^3$. However, in cases where the location parameter is fixed (e.g., $x_0=0$ for the classical Weibull distribution), we can uniquely identify the scale and shape parameters.

When we consider quantiles as observations, we investigate the impact of the number of quantiles, denoted as $M$, on the posterior distribution. Specifically, we choose the quantile values $(p_j)_{j=1,\dots,M}$ such that $p_j= \frac{j}{M+1}$ for $j=1,\dots,M$. For example, when $M=3$, we have $(p_1,p_2,p_3) = (0.25,0.5,0.75)$, and for $M=9$, we obtain the nine deciles.

As observed in Figure \ref{fig:weibull_theoretical}, it becomes apparent that using only two quantiles is insufficient to capture all three parameters accurately. However, with a minimum of three quantiles, we can successfully identify all the parameters of the Weibull distribution. Additionally, increasing the number of quantiles leads to a posterior distribution that closely aligns with the theoretical parameters, indicating improved estimation precision.
\end{appendices}

\end{document}